# Reliable and Repeatable Transit Through Cislunar Space Using the 2:1 Resonant Spatial Orbit Family

Andrew Binder* and David Arnas[†]

*Purdue University, West Lafayette, IN 47907, USA*

This work focuses on the identification of reliable and repeatable spatial (three-dimensional) trajectories that link the Earth and the Moon. For this purpose, this paper aims to extend the 2:1 resonant prograde family and 2:1 resonant retrograde family to three dimensions and to introduce spatial orbits that are not currently present in the literature. These orbits, named the 2:1 resonant spatial family, bifurcate from the two-dimensional families and smoothly transition between them in phase space. The stability properties of this new family of resonant orbits are discussed, and, interestingly, this family includes marginally stable members. Furthermore, this new family of orbits is applied to several engineering problems in the Earth-Moon system. First, this paper selects an appropriate member of the 2:1 resonant spatial family on the basis of its stability properties and relationships with other multibody orbits in the regime. Next, this work combines this trajectory with momentum exchange tethers to transit payloads throughout the system in a reliable and repeatable fashion. Finally, this paper studies the process of aborting a catch and related recovery opportunities.

## I. Introduction

The cislunar environment has been the subject of considerable attention [1–4] in recent years, as illustrated by the initial successes of the *Artemis I* [5] and *CAPSTONE* [6] missions. This has shown that the efficient and reliable traversal of this regime is paramount to the success of future missions, the continued success of cislunar operations, and the development of the cislunar economy. Spacecraft trajectories through cislunar space are governed by the combined gravitational pulls of the Earth and Moon on the spacecraft, generating a resultant dynamic governed by the three-body problem. However, the general three-body problem is known to have no analytical solution but for a very small number of particular cases. These particular solutions are not applicable to the Earth-Moon problem or cislunar space, and therefore numerical approaches are required to study and understand the dynamics of this system. In particular, one of the most well-known and used models to represent this system is the so-called Circular Restricted Three-Body Problem (CR3BP), where we assume Newtonian gravity [7], circular orbits for each primary, and a small third body that is free to move throughout the evolving system and that does not affect the dynamics of the two primaries. The CR3BP, although a simple model, provides good insight into the dynamics of the system due to its relative simplicity, while still providing an acceptable first-order solution for higher-fidelity methods. As such, this work focuses on the use of the CR3BP to provide a better understanding of the dynamics of this regime and, specifically, the description of reliable and repeatable transit orbits in cislunar space. In this paper, this is done through the use of resonant orbit families in the CR3BP.

---

*Graduate Researcher, School of Aeronautics and Astronautics, binder1@purdue.edu.

[†]Assistant Professor, School of Aeronautics and Astronautics, darnas@purdue.edu.

In general, resonant orbits and their associated families serve an important purpose in the literature. Described first by References [8, 9], the planar families discussed here can be shown to span cislunar and circumlunar space. Further work in the 1990s by de Almeida Prado and Broucke [10, 11] developed the study of these resonant orbits and transfers between them in the planar case. More recent work, including works by Vaquero and Howell [12], Vaquero [13], and Gupta [14] have levied dynamical systems theory and computational power to discover transfers between such orbits in the three-dimensional (e.g. 'spatial') case. These families have proven to be practical in many applications, such as in the field of space situational awareness. For example, Gupta et al. [15, 16] demonstrated that resonant orbit families can be used for long-term surveillance of the cislunar region, the circumlunar region, and surrounding regions at repeated intervals.

However, no previous resonant orbit family discussed in the literature can span three-dimensional cislunar space, pass nearby to the Earth and Moon, and repeat this dynamic on a frequent basis. This paper seeks to identify and describe a family of periodic trajectories, which we refer to as the *transit* family, which satisfies these desired properties. This family spans cislunar space in the three-dimensional CR3BP and is shown to be related to a set of periodic resonant orbits (ROs) described before and analyzed in the literature [8, 9] for the simpler planar problem. Beyond the description of the *transit* family, this paper presents a case study for trajectory design that uses this transit family to achieve transfer trajectories from Low Earth Orbit (LEO) to cislunar halo orbits. These transfers repeat frequently when compared to typical low-energy transfers. Finally, this set of transfer trajectories is also shown to be safe for operational complications, such as a missed injection into a halo.

This paper is organized into six sections. First, in the Introduction (Section I), the motivation for the problem is presented, as well as previous related work. In Section II, the CR3BP is summarized and a discussion of resonant orbits in the conic problem is presented. In Section III, the arsenal of numerical tools used in this work is described. This includes shooting methods, state transition matrices, Floquet theory, bifurcations, and the use of the Broucke stability diagram (BSD) to identify bifurcations. In Section IV, the newly discovered 2:1 resonant spatial family of periodic orbits (collectively called the *transit family*) is presented. This includes a discussion on their origin from related planar RO counterparts and an analysis of the properties of individual orbits of the family. In Section V, a member of the newly introduced family is selected to demonstrate the applicability of the proposed family of orbits and design methodology to cislunar trajectory design. In Section VI, some operational concerns are addressed. That is, nearly homoclinic connections back to the chosen member of the *transit family* are identified and discussed as a recovery strategy in the context of the other content in Section V. Finally, Section VII includes a summary of the findings presented in this paper as well as their implications for current and future cislunar missions.

## II. Description of the model

### A. Coordinate frames, nondimensionalization, and equations of motion for $m_3$

We non-dimensionalize by first defining the three quantities $l^*$ (the characteristic length), $m^*$ (the characteristic mass), and $t^*$ (the characteristic time):

$$l^* = \|\mathbf{R}_1\| + \|\mathbf{R}_2\| \qquad m^* = m_1 + m_2 \qquad t^* = \sqrt{\frac{(l^*)^3}{Gm^*}} = \sqrt{\frac{(l^*)^3}{\mu_1 + \mu_2}} \qquad (1)$$

where $\mathbf{R}_1$ and $\mathbf{R}_2$ are the positions of the two primaries with respect to their barycenter, $m_1$ and $m_2$ are their masses, and $G$ is the gravitational constant.

With these definitions, the dimensional mean motion of the two primaries is as follows:

$$\mathcal{N} = \sqrt{\frac{Gm^*}{(l^*)^3}} \qquad (2)$$

where $Gm^* = Gm_1 + Gm_2 = \mu_1 + \mu_2$ is the sum of the dimensional $\mu_S$ values of the primaries. This allows defining $t^*$ to cause the non-dimensional mean motion $n = \mathcal{N}t^* = 1$. Additional non-dimensionalizations for the masses $m_1$ and $m_2$, the position and velocity vectors $\mathbf{R}$ and $\mathbf{V}$, and time $t$, are given by the following:

$$\mathbf{r} = \frac{\mathbf{R}}{l^*}, \qquad \mathbf{v} = \frac{\mathbf{V}t^*}{l^*} = \frac{\mathbf{V}}{v^*}, \qquad \tau = \frac{t}{t^*}, \qquad \mu = \frac{m_2}{m^*} = 1 - \frac{m_1}{m^*} \qquad (3)$$

In particular, for the Earth-Moon system in this paper, the CR3BP constants used are provided in Table 1, where [n.d.] represents a non-dimensional quantity.

#### Table 1    Earth-Moon characteristic quantities

| $\mu_1$, Earth [km³/s²] | $\mu_2$, Moon [km³/s²] | $l^*$ [km] | $t^*$ [s] (derived) | $\mu$ [n.d.] (derived, trunc. to 8 s.f.) |
|---|---|---|---|---|
| 398600.4415 | 4902.8005821478 | 384400 | 375190.26 | 0.012150585 |

We define the axis about which the two primaries rotate as $\mathbf{u}_z$, and the angle subtended by their rotation over time is $\theta = n(t - t_0) = \tau - \tau_0$, where $\tau_0$ is the epoch we set in our model. As a result of our non-dimensionalization, when written in the rotating frame $\mathbf{R}_1 = -\mu\,\mathbf{u}_x + 0\,\mathbf{u}_y + 0\,\mathbf{u}_z$ [n.d.] and $\mathbf{R}_2 = (1 - \mu)\,\mathbf{u}_x + 0\,\mathbf{u}_y + 0\,\mathbf{u}_z$ [n.d.]. Figure 1 depicts the relationships between the frames and the masses pictorially. For the sake of brevity for the remainder of this text, all vectors will be written in non-dimensional coordinates with respect to the rotating frame $\{\mathbf{u}_x, \mathbf{u}_y, \mathbf{u}_z\}$ unless otherwise specified.

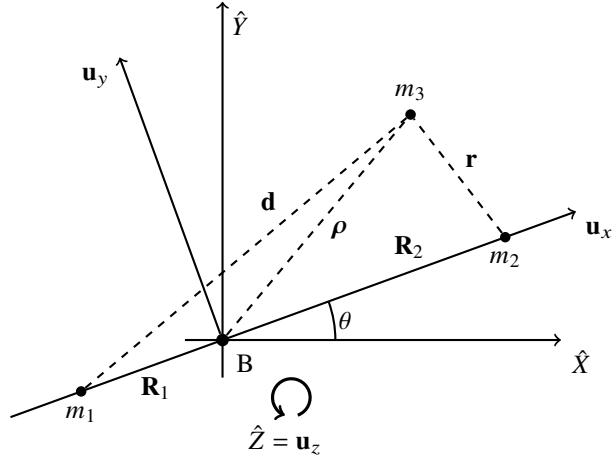

**Fig. 1    Depiction of the coordinate frames used and vector definitions**

Our third mass $m_3$ (also modeled as a point mass $m_3 \ll m_1, m_2$) is at a position $\boldsymbol{\rho} = x\,\mathbf{u}_x + y\,\mathbf{u}_y + z\,\mathbf{u}_z$ [n.d.] and a velocity $\dot{\boldsymbol{\rho}} = \dot{x}\,\mathbf{u}_x + \dot{y}\,\mathbf{u}_y + \dot{z}\,\mathbf{u}_z$ [n.d.] with respect to the barycenter $B$. Its distance from $m_1$ is $\mathbf{d} = (x + \mu)\,\mathbf{u}_x + y\,\mathbf{u}_y + z\,\mathbf{u}_z$ [n.d.] and its distance from $m_2$ is $\mathbf{r} = (x + 1 - \mu)\,\mathbf{u}_x + y\,\mathbf{u}_y + z\,\mathbf{u}_z$ [n.d.]. Expressed in the rotating frame, the dynamics admits the scalar equations of motion (EOM) [17]:

$$
\begin{aligned}
\ddot{x} - 2\dot{y} &= \frac{\partial U}{\partial x} \\
\ddot{y} + 2\dot{x} &= \frac{\partial U}{\partial y} \\
\ddot{z} &= \frac{\partial U}{\partial z}
\end{aligned}
\tag{4}
$$

where $U(\boldsymbol{\rho})$ is the pseudopotential energy:

$$
U(\rho) = \frac{1}{2}(x^2 + y^2) + \frac{1 - \mu}{d} + \frac{\mu}{r}; \qquad d = \|\,\mathbf{d}\,\| = \sqrt{(x + \mu)^2 + y^2 + z^2}; \qquad r = \|\,\mathbf{r}\,\| = \sqrt{(x + 1 - \mu)^2 + y^2 + z^2}
\tag{5}
$$

and $d$ and $r$ are the magnitudes of the position vectors of $m_3$ relative to the two primaries, respectively. Integration of these EOM are done in the *MATLAB ODE Suite* [18].

### B. Equilibrium points and the Jacobi Constant

These autonomous CR3BP EOM have exactly five equilibrium points (called "libration points" or "Lagrange points") $L_1$-$L_5$. These five points ($L_1$-$L_3$ colinear and $L_4$-$L_5$ equilateral) are the precise locations where the gravitational influences of the primaries and the virtual rotational forces cancel. A diagram of these points and their locations relative to the primaries can be found in Figure 2.

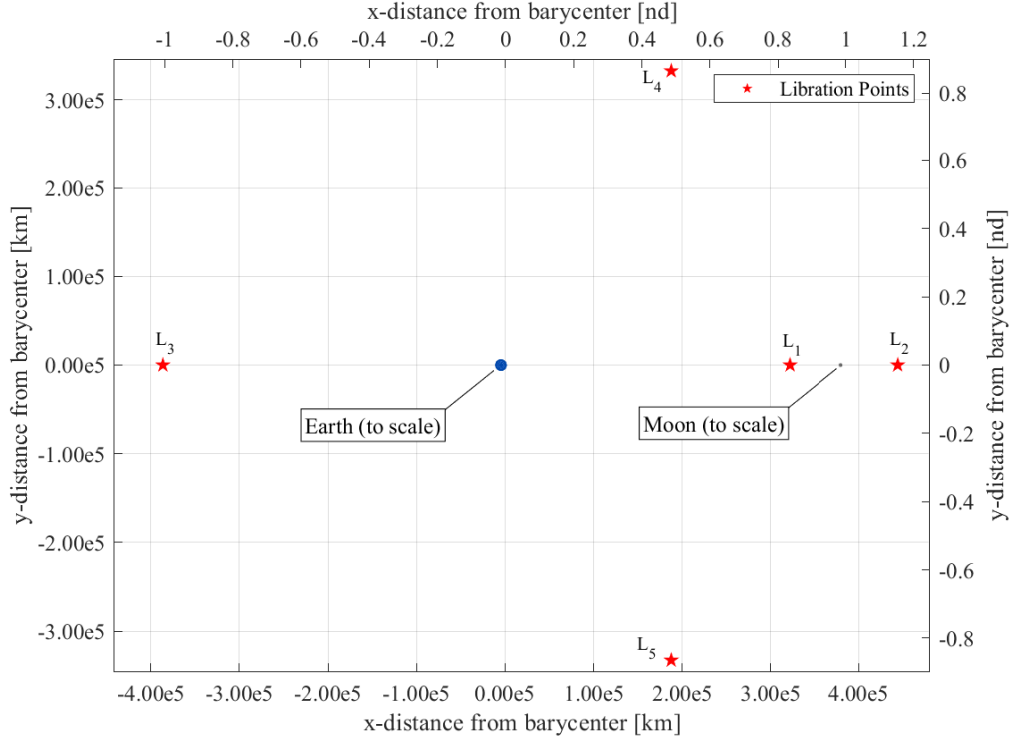

**Fig. 2  The locations of $L_1$-$L_5$ in rotating coordinates, dimensional and nondimensional units**

In the rotating frame, there is a single known analytical integral of the motion, Jacobi Constant ($JC$) (an energy-like quantity). It is defined as follows:

$$JC(\boldsymbol{\rho}, \dot{\boldsymbol{\rho}}) = 2U(\boldsymbol{\rho}) - \dot{x}^2 - \dot{y}^2 - \dot{z}^2 = 2U(\boldsymbol{\rho}) - \| \dot{\boldsymbol{\rho}} \|^2 = 2U - v^2 \tag{6}$$

$JC$ as a metric for assessing trajectories is useful in a number of ways, including the following four. First, as a constant of motion, $JC$ reduces the dimension of solutions $x(t)$ to codim(1) curves under the dynamics. Second, it can be used as a medium to assess the accumulated error in simulations. Third, $JC$ can be used to create boundaries that partition the phase space into *permissible regions* and *forbidden regions*. This partition (separated in the plane by the Zero Velocity Curve (ZVC) and in three dimensions by a surface called the Zero Velocity Surface (ZVS) [17]) is helpful in providing information about possible motions and trajectories throughout the system. Finally, $JC$ is helpful as a means of measuring the optimality of an impulsive transfer between two trajectories. Rearranging the $JC$ formula for the velocity, we get the following.

$$v = \sqrt{2U - JC}$$

At the same point in space $\boldsymbol{\rho}$ (where $U$ has a specified value), the smallest $\Delta v$ possible [19] to impulsively insert from one trajectory into the other is $\Delta v = v_2 - v_1$, or:

$$\Delta v = v_2 - v_1 = \sqrt{2U - JC_2} - \sqrt{2U - JC_1} \qquad (7)$$

### C. Mean-motion resonant orbits

*1. Resonant orbits in the conic problem*

Resonance appears as a natural phenomenon often in the orbits of bodies throughout the Solar System. For this text, we define resonance to be some rational fraction of instantaneous orbital periods $p/q$ (also written $p : q$, alternatively deemed *mean-motion orbital resonance*). $p$ and $q$ are the set of natural co-prime numbers $(p, q) \in \mathbb{N}$ where $\gcd(p, q) = 1$ that satisfy the following relation:

$$p\ \mathbb{P}_1 - q\ \mathbb{P}_2 = 0 \qquad (8)$$

where $\mathbb{P}_1$ is the period of the first body and $\mathbb{P}_2$ is the period of the second.

In this work, we refer to the period of the spacecraft ($m_3$) as $\mathbb{P}_1$ and the period of the smaller primary ($m_2$) about the larger primary ($m_1$) as $\mathbb{P}_2$. In this text, we often refer to *sidereal* and *synodic* periods. When we discuss *sidereal* periods, periodicity is measured with respect to an inertial frame. Similarly for *synodic* periods (especially when discussing the *synodic* period of $m_3$), periodicity is measured with respect to position relative to the Moon. $\mathbb{P}_2$ is specifically taken to be the sidereal period of the Moon orbiting about its mutual barycenter with Earth. In the Keplerian case, the orbital period is calculated classically:

$$\mathbb{P}_2 = 2\pi \sqrt{\frac{a^3}{\mu_1 + \mu_2}} \qquad (9)$$

where $a$ is the orbit's semi-major axis, $\mu_1$ is the standard gravitational parameter of one orbiting body and $\mu_2$ is the standard gravitational parameter of the other. Note that, when calculating the orbital period of $m_3$ about $m_1$ or $m_2$, the equation reduces to approximately the following:

$$\mathbb{P}_1 = 2\pi \sqrt{\frac{a^3}{\mu_3 + \mu_{1,2}}} \approx 2\pi \sqrt{\frac{a^3}{\mu_{1,2}}}$$

When discussing orbital resonance of this kind, we define the term *simple resonance* to mean some $p, q$ that are "small" and create a "simple" fraction. Although imprecise, for the sake of discussion, it is helpful to clarify the taxonomy with a few examples at the extremes. Under this definition, the resonances $2 : 1$ and $3 : 8$ would be considered *simple*, whereas the resonance $582 : 233$ would not.

## 2. Generalization of ROs in the CR3BP

Resonant orbits of a simple $p : q$ exist and are relatively easy to produce in the Keplerian model. What is less straightforward is finding an analogue of these orbits in the CR3BP. Previous studies by Vaquero [13] and Gupta et al. [15] have produced and studied a wide range of these analogs (planar and three-dimensional) in the CR3BP. Methods (such as those used by Vaquero, Gupta et al. and others) exist to numerically correct 'good enough' initial conditions such that, when simulated, they meet some condition at some specified time. One such class of methods (derived from the calculus of variations) is called single-shooting methods and is described more thoroughly in Section III.A. These methods (used widely in the fields of dynamical systems and numerical astrodynamics) can target many end conditions, including periodicity.

# III. Numerical Methodology

In nonlinear dynamical systems, obtaining analytical solutions is often challenging or impossible. Approximations, such as linearizing the dynamics, can provide insight into the problem but often lead to inaccurate quantitative predictions of behavior. Therefore, employing *numerical methods* to generate point solutions to study the broad behaviors of dynamical systems becomes crucial to effective design and understanding. When combined with dynamical systems theory, these methods enable a thorough exploration of a given problem, helping designers make informed decisions.

## A. Shooting methods, the state transition matrix (STM) and numerical continuation

Elaborating on the discussion at the end of Section II.C.2, ROs in the conic problem can be generalized to the CR3BP. By using single-shooting methods, the state transition matrix, and numerical continuation, it is possible to perform a numerical homotopy that smoothly deforms the topology of solutions from Keplerian dynamics into solutions in the CR3BP. Starting with Keplerian dynamics (where effectively the non-dimensional mass parameter $\mu$ is zero), we can vary the value of $\mu$ to iteratively approach CR3BP dynamics, converging onto periodic solutions at each step in $\mu$ using shooting methods. In this text, we perform this homotopy to transform one member of each planar resonant orbit family from Keplerian to CR3BP dynamics. Once one member is in hand, an additional numerical continuation scheme can be used to find other members of the family.

Shooting methods are a category of numerical techniques designed to solve boundary value problems. In nonlinear dynamical systems, finding solutions to these boundary value problems can be difficult, so it can be advantageous to treat the problem as a root-finding problem (a numerical challenge studied extensively in the literature). In this work, we focus on the simplest class of shooting methods, namely *single-shooting methods*. Consider the state $\mathbf{X}_f$ at time $\tau_2$ and let $\mathbf{F}$ be a $m \times 1$ column vector with elements $f = f(\mathbf{X}_f, \tau_2) \in \mathbb{R}$, where each element of this vector represents a constraint placed on the state at that time. All constraints are "satisfied" when $\| \mathbf{F} \| = 0$. Assuming that we can only control the state $\mathbf{X}_0$ at a different time $\tau_1$, we need to find an efficient way to determine $\mathbf{X}_0(\tau_1)$ that results in $\| \mathbf{F} \| = 0$. An effective approach is to utilize the *state transition matrix (STM)* $\Phi(\tau_1, \tau_2)$ [20, pp.8-10], which is a linear stroboscopic map defined between variations about states at the starting time $\tau_1$ and another time $\tau_2$. The $6 \times 6$ STM $\Phi$ with elements $\phi_{ij} \in \mathbb{R}$ linearly estimates the impact of modifying the state at $\tau_1$ on the state at $\tau_2$. Each element of $\Phi$ represents the partial derivative of a specific element of the final state with respect to a particular element of the initial state. Specifically, if $\mathbf{X}_0(\tau_1) = (x_0,\ y_0,\ z_0,\ \dot{x}_0,\ \dot{y}_0,\ \dot{z}_0)^T$ and $\mathbf{X}_f(\tau_2) = (x_f,\ y_f,\ z_f,\ \dot{x}_f,\ \dot{y}_f,\ \dot{z}_f)^T$:

$$
\Phi = \begin{bmatrix}
\frac{\partial x_f}{\partial x_0} & \frac{\partial x_f}{\partial y_0} & \frac{\partial x_f}{\partial z_0} & \frac{\partial x_f}{\partial \dot{x}_0} & \frac{\partial x_f}{\partial \dot{y}_0} & \frac{\partial x_f}{\partial \dot{z}_0} \\[6pt]
\frac{\partial y_f}{\partial x_0} & \frac{\partial y_f}{\partial y_0} & \frac{\partial y_f}{\partial z_0} & \frac{\partial y_f}{\partial \dot{x}_0} & \frac{\partial y_f}{\partial \dot{y}_0} & \frac{\partial y_f}{\partial \dot{z}_0} \\[6pt]
\frac{\partial z_f}{\partial x_0} & \frac{\partial z_f}{\partial y_0} & \frac{\partial z_f}{\partial z_0} & \frac{\partial z_f}{\partial \dot{x}_0} & \frac{\partial z_f}{\partial \dot{y}_0} & \frac{\partial z_f}{\partial \dot{z}_0} \\[6pt]
\frac{\partial \dot{x}_f}{\partial x_0} & \frac{\partial \dot{x}_f}{\partial y_0} & \frac{\partial \dot{x}_f}{\partial z_0} & \frac{\partial \dot{x}_f}{\partial \dot{x}_0} & \frac{\partial \dot{x}_f}{\partial \dot{y}_0} & \frac{\partial \dot{x}_f}{\partial \dot{z}_0} \\[6pt]
\frac{\partial \dot{y}_f}{\partial x_0} & \frac{\partial \dot{y}_f}{\partial y_0} & \frac{\partial \dot{y}_f}{\partial z_0} & \frac{\partial \dot{y}_f}{\partial \dot{x}_0} & \frac{\partial \dot{y}_f}{\partial \dot{y}_0} & \frac{\partial \dot{y}_f}{\partial \dot{z}_0} \\[6pt]
\frac{\partial \dot{z}_f}{\partial x_0} & \frac{\partial \dot{z}_f}{\partial y_0} & \frac{\partial \dot{z}_f}{\partial z_0} & \frac{\partial \dot{z}_f}{\partial \dot{x}_0} & \frac{\partial \dot{z}_f}{\partial \dot{y}_0} & \frac{\partial \dot{z}_f}{\partial \dot{z}_0}
\end{bmatrix}
\tag{10}
$$

$\Phi(\tau, \tau_1)$ (the matrix that maps the states at time $\tau_1$ to some future time $\tau$) evolves according to the first-order matrix differential equation.

$$
\dot{\Phi}(\tau, \tau_1) = A(\tau)\Phi(\tau, \tau_1), \quad \Phi(\tau_1, \tau_1) = I_{6 \times 6}
\tag{11}
$$

where coefficients $A(\tau)$ are found by linearizing the nonlinear differential equations to produce *variational equations*. In practice, $\Phi$ is integrated alongside the state EOM. To solve our class of two-point boundary value problems, we select a set of *design variables* $\mathbf{Q}$ where $\mathbf{Q}$ is a $n \times 1$ column vector and each element $q$ of $\mathbf{Q}$ is a function of the initial state and time $q = f(\mathbf{X}_0, \tau_1)$. For this paper, we only consider the case where $m = n$ or, equivalently, $\mathbf{F}$ and $\mathbf{Q}$ have the same length. To drive $\| \mathbf{F} \| = 0$, we want a specific value of the design variables $\mathbf{Q}^*$ that satisfies our constraints. Performing a Taylor expansion of $\mathbf{F}(\mathbf{Q})$ about our initial guess $\mathbf{Q}_k$, we can write $\mathbf{F}_k(\mathbf{Q}) = DF_k(\mathbf{Q}_k)(\mathbf{Q} - \mathbf{Q}_k)$. Here, $DF$ is the Jacobian matrix of $F$ where the elements of $DF$ are $\frac{\partial \mathbf{F}(\mathbf{Q}_k)}{\partial \mathbf{Q}_k}$, or the rate of change of elements of $\mathbf{F}$ as the elements of $\mathbf{Q}$ are modified. In general, elements of $DF$ are functions of elements of each state, each time $\tau$ and elements of the STM. We can write the set of design variables that satisfies our constraints as $\mathbf{Q}^* = \mathbf{Q}_k + \Delta\mathbf{Q}_k$ where $\mathbf{Q}_k$ is the current value of our design variables and $\Delta\mathbf{Q}_k$ is some change in our design variables. Combining these two equations, we can write an update equation to iteratively approach $\mathbf{F}(\mathbf{Q}^*) = \mathbf{0}$:

$$
\begin{aligned}
\mathbf{Q}_{k+1} &= \mathbf{Q}_k + \Delta\mathbf{Q}_k \\
\Delta\mathbf{Q}_k &= -DF_k^{-1}\mathbf{F}_k
\end{aligned}
\tag{12}
$$

Each update of $\mathbf{Q}_k$ requires a reintegration of the equations of motion and a subsequent reevaluation of the error vector

$\mathbf{F}_k(\mathbf{Q})$. With a 'good' starting $\mathbf{Q}_0$, $\lim_{k\to\infty}\mathbf{Q}_k \to \mathbf{Q}^*$. Using shooting methods, we can converge on many conditions, including periodicity. In dynamical systems, these periodic solutions are called *orbits*. Such *orbits* are often members of *families* of similar orbits. Specifically in the CR3BP once a first periodic orbit is found, it is always possible in the CR3BP to find some 'nearby' periodic orbits that are members of the same family. The members of such families in the CR3BP are packed infinitely densely in the family. More precisely, given two members of the same family, it is always possible to find some additional member that is an interpolation between the original two members. A simple strategy to construct a family of orbits is called natural parameter continuation. Once some $\mathbf{Q}^*$ is found for one periodic orbit, it is possible to vary parameters of the problem or solutions unrelated to $\mathbf{Q}^*$ by a small amount. This variation can lead to new solution sets of design variables $\mathbf{Q}^*$ and this process can be repeated.

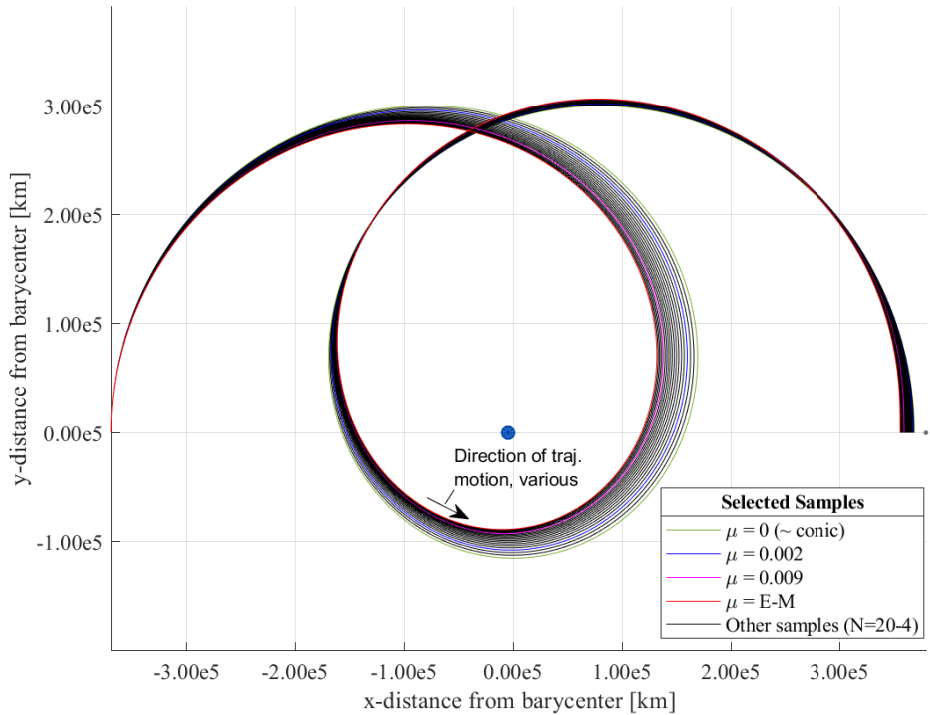

**Fig. 3   A natural parameter continuation in $\mu$ [nd] to generalize a 2:1 retrograde family member in the CR3BP**

Performing a continuation in the nondimensional mass parameter $\mu$ was first performed by Arenstorf [8] and recently by Gupta et al. [15]. Both sources produce ROs families generalized to the CR3BP by employing numerical continuation methods. As explained at the beginning of this section, one homotopic process for one Keplerian orbit is only the first step. Finding this first periodic solution is visualized in Figure 3. Once a single periodic solution is found, it is possible to employ other parameter continuations to generate families of periodic orbits. Specifically of interest in this paper, Gupta et al. produces families of orbits related to the 2:1 prograde and retrograde resonant orbits, reproduced and discussed in Section IV.A. In this work, we extend these families by performing a natural parameter continuation to transition this family into a three-dimensional space. To do so, we specifically seek departures and returns to the xz plane that crosses the plane in a perpendicular manner. This process, used by Gupta et al. [15] for the planar case and demonstrated in Figure 3, is equally applicable to the three-dimensional case. Throughout this paper, we use this method and the Mirror Theorem (described by Roy and Ovenden [21]) to produce periodic orbits.

## B. The monodromy matrix and Floquet multipliers

Another application of the STM is to analyze the stability properties of periodic orbits and their families. The monodromy matrix $\mathcal{M} = \Phi(\tau_1 + T, \tau_1)$ is defined as the STM integrated alongside a precisely periodic orbit for one orbital period $T$. Often in this paper, we take $T$ to be the synodic period of a given periodic orbit. The eigenvalues of this monodromy matrix (also called Floquet multipliers) can be used to assess the stability properties of the periodic orbit associated with this monodromy matrix [20, pp.22-32].

## C. Bifurcations along orbit families and the Broucke stability diagram (BSD)

Members of families of orbits are considered one-dimensional equilibrium solutions, and the families of these solutions when considered as a whole have numerous interesting properties. One of interest to us is the stability properties of these equilibria and how they change over the extent of a full family. Scanning through the members of a family, it is possible to find specific locations in the family where stability changes occur. A stability change is defined as a change in the character or composition of the eigenvalues of the monodromy matrix. At these stability changes, multiple families of orbits can coincide [20, pp.156-165]. A useful tool for finding these stability changes is the Broucke stability diagram (BSD), introduced by Broucke [22], generalized by Howard and MacKay [23] and developed in depth for the CR3BP by Campbell [24] and Grebow [25]. An example BSD is visualized in Figure 4, shown without data from a family plotted on the graph. This example graph contextualizes future analysis by labeling each region of stability and identifying the transitions between them. Data analysis using a BSD is performed later in Figure 7.

Generally, members of a family are found and their monodromy matrices $\mathcal{M}_i$ are considered. For each $\mathcal{M}_i$, six eigenvalues $\lambda_j$ can be obtained. However, these eigenvalues are not independent of each other. In a purely periodic orbit in the CR3BP, two Floquet multipliers will always be $+1 + 0i$ and, since the elements of $\mathcal{M} \in \mathbb{R}$ and the system is governed by a Hamiltonian, the four remaining eigenvalues come in reciprocal pairs $\lambda$ & $1/\lambda$. Broucke [22] discusses these properties and presents his stability coefficients $a_1$, $a_2$ that consolidate the pairs of non-unity eigenvalues $(\lambda_1, 1/\lambda_1)$ and $(\lambda_2, 1/\lambda_2)$. As shown in Equation (13) using the method of Leverrier [26, Appendix 6], these stability coefficients can be computed more efficiently and accurately using the *trace Tr()* of the monodromy matrix. This precludes the need to solve the eigenvalue problem, a numerical challenge that can suffer from inaccuracy.

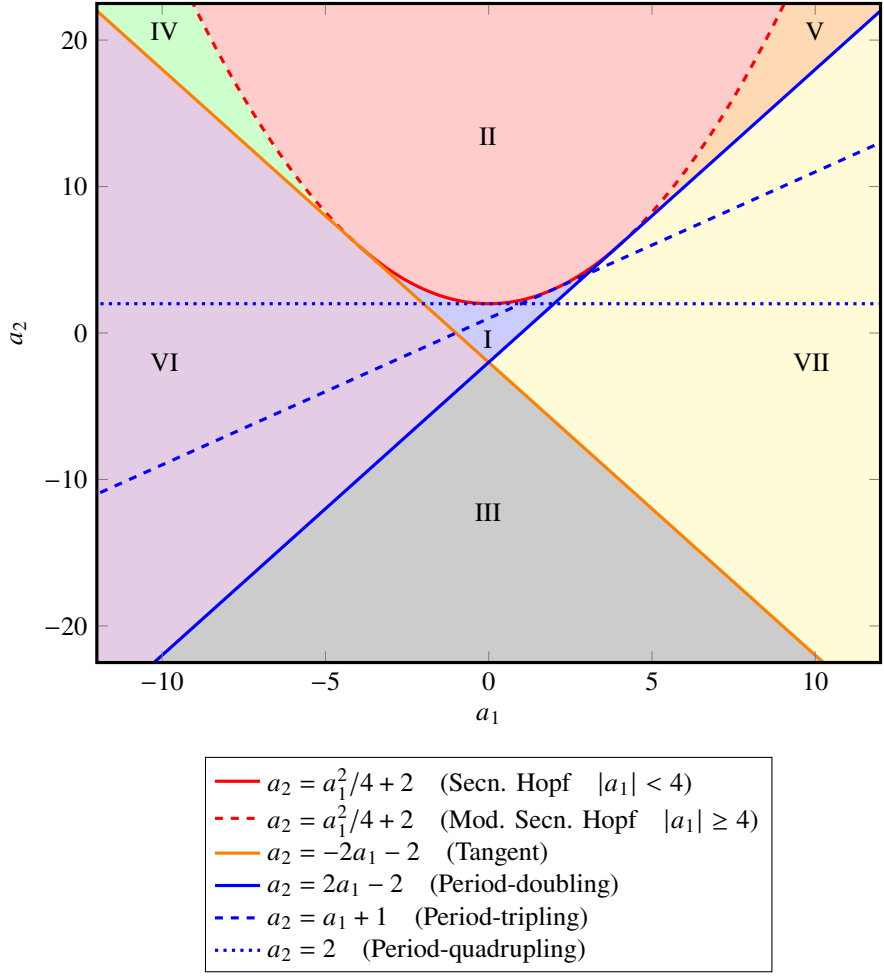

**Fig. 4   A Broucke stability diagram with lines of bifurcation [22]**

$$
\begin{aligned}
a_1 &= -\left(\lambda_1 + \frac{1}{\lambda_1} + \lambda_2 + \frac{1}{\lambda_2}\right) = 2 - Tr(\mathcal{M}) \\
a_2 &= 2 + \left(\lambda_1 + \frac{1}{\lambda_1}\right)\left(\lambda_2 + \frac{1}{\lambda_2}\right) = \frac{1}{2}\left(a_1^2 - \left(\lambda_1^2 + \frac{1}{\lambda_1^2} + \lambda_2^2 + \frac{1}{\lambda_2^2}\right)\right) = \frac{1}{2}\left(a_1^2 + 2 - Tr(\mathcal{M}^2)\right)
\end{aligned}
\tag{13}
$$

Each member of the family receives a pair of coordinates $(a_1, a_2)$ and these coordinates are plotted as a discontinuous set of points on the BSD. As the discontinuous set of points $(a_1, a_2)$ trace an arc on the BSD, we look for places where this arc crosses one of the many lines visualized above. The equations of these lines are derived by Broucke and Campbell with some equations included in the legend of Figure 4. It is at these crossings that we identify bifurcations in the family analyzed. Of interest to us are *period-doubling bifurcations*. At such bifurcations, two families of periodic solutions intersect at that bifurcating point, with members of one family after the bifurcation having twice the period of the original family at that point [20, p. 158]. As discussed in the following section, this kind of *period-doubling* *bifurcation* can be found in both planar families.

## IV. The 2:1-resonant planar and spatial families

### A. Unique properties of the planar Earth-Moon 2:1 resonant families

First described by Arenstorf [8] and Broucke [9] respectively, and later in considerable detail by Gupta [14], the 2:1 prograde and retrograde planar families (visualized in Figure 5) have a number of properties relevant to this text *. Of the "simple" resonant orbit families studied previously in the literature [12–16], R2:1-P and R2:1-R are the only known families that: 1) span cislunar space[†], 2) potentially possess a spatial variant that does the same, and 3) pass 'close' to the surfaces of both the Earth and Moon. The spatial (e.g. three-dimensional) variants of R2:1-P and R2:1-R are found in this paper to exist, are found to emerge from period doubling bifurcations and, surprisingly, are found to be the same family. This is described in detail in the following subsections.

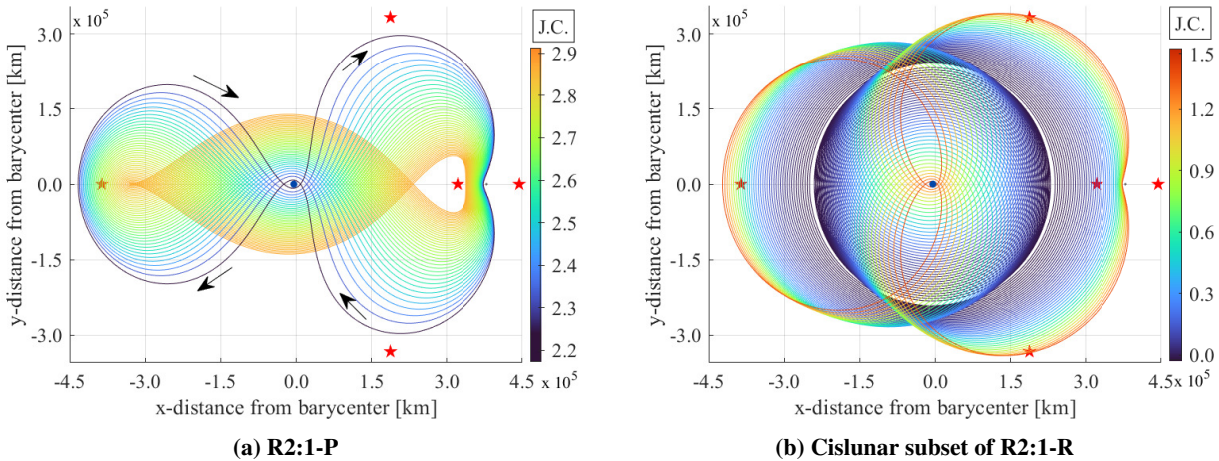

(a) R2:1-P                        (b) Cislunar subset of R2:1-R

**Fig. 5    The families associated with the planar 2:1-resonant orbits
(arrows are general direction of orbital motion)**

### B. The spatial Earth-Moon 2:1 resonant family

In Figure 6, we present the Earth-Moon 2:1 resonant spatial family (R2:1-S). The spatial family links the planar orbit families R2:1-P and R2:1-R. More precisely, R2:1-S bifurcates at one end from R2:1-P, bifurcates at the other end from R2:1-R, and these two planar bifurcating orbits are linked by a smooth and continuous set of orbits in three dimensions. This spatial family of orbits meets all three desired requirements, as outlined in Section I. First, it spans cislunar space in three dimensions and is not purely planar. Second, it passes closely to the surfaces of the Earth and

---

*Colloquially, we will refer to these families as "R2:1-P" (for the prograde 2:1 resonant family) and "R2:1-R" (for the retrograde 2:1 resonant family) for brevity. Later, we will introduce "R2:1-S" (for the spatial (3D) 2:1 resonant family).

[†]R2:1-R as shown in Figure 5 is a subset of a larger family of retrograde trajectories that extend past the Moon and into circumlunar space. This circumlunar segment of the family is not of interest to this paper and, as such, is not displayed.

Moon. Third and finally, it does so on a frequent basis, with each pass by Earth separated in time by approximately half of a lunar sidereal period (roughly 14 days) and with each pass by the Moon separated in time by approximately one lunar sidereal period (roughly 28 days).

R2:1-S emerges from both R2:1-P and R2:1-R at period-doubling bifurcations. The locations of these bifurcations (and others in R2:1-P and R2:1-R) are shown plotted on a BSD in Figure 7. Although a number of bifurcations are present in R2:1-R, the bifurcation plotted as a red dot corresponds to the point of origin of R2:1-S. The set of orbits within R2:1-P only has one bifurcation: the period-doubling bifurcation of R2:1-S's origin, also indicated with a red dot. We refer to the member at the location of each bifurcation as the 'bifurcating orbit'. To provide further insight into the location of each bifurcation, each subfigure of Figure 5 was recreated with all members plotted in a desaturated color, as shown in Figure 8. The bifurcating orbits are overlayed, colored in more saturated blue or red (for R2:1-P and R2:1-R, respectively).

To report initial conditions that result in these spatial periodic orbits, we define a finite spatial surface of section "$A$" that rotates with the Earth and Moon and report our measurements of this family at intersections with that surface. $A$ is the plane in position space that, in rotating x-coordinates, spans between Earth-Moon $L_1$ and the Moon's center of mass. The surface of section is also defined to measure positive rotating z-coordinates $z > 0$ (to avoid duplicate measurement of family members). The entire surface of section lies within the rotating $y = 0$ plane. Although R2:1-S is a continuous and smooth set of orbits, it is convenient to partition the family into subsets that share qualitatively similar behaviors. Presented are several hodographs that describe the family. In Figure 11 (which shows the stability index of each member of the family), there is a specific member of the family that has the highest stability index in the family. In this text, stability index $\nu$ is calculated using the eigenvalue of maximum absolute value $\lambda_{max}$ as:

$$\nu = \frac{1}{2}\left(|\lambda_{max}| + \frac{1}{|\lambda_{max}|}\right) \tag{14}$$

It is near this member that the two similar semicircular close-approach arcs of the family (of Figure 9(a)) intersect, and it is at this member that we perform our partition. We refer to members that are more 'prograde' than the partitioning member as R2:1-S1 and refer to members that are more 'retrograde' than the partitioning member as R2:1-S2. This partition (and the location of the partitioning, or 'splitting' member), is reflected in Figure 9 (mentioned previously) and also in Figure 10. For every crossing shown in Figure 9 with $\dot{y} < 0$, there is an identical crossing with $z \to -z$ also with $\dot{y} < 0$ (e.g. the family is symmetric by reflection across the xy-plane). The velocity $\dot{y}$ at each crossing is the same and each crossing occurs perpendicular to $A$ (for example, $\dot{x} = \dot{z} = 0$ at crossings of $A$).

Several interesting features of the family deserve attention. First, note that the members of R2:1-S pass at varying distances over the surface of the Moon. Also note that the orbits pass the Moon at a wide range of relative velocities. The net effect of these two observations in tandem is displayed in Figure 10(b), a figure that describes that family members

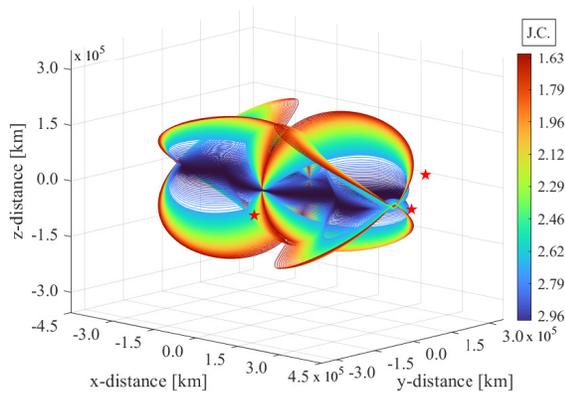

**(a) The prograde-adjacent half: R2:1-S1**

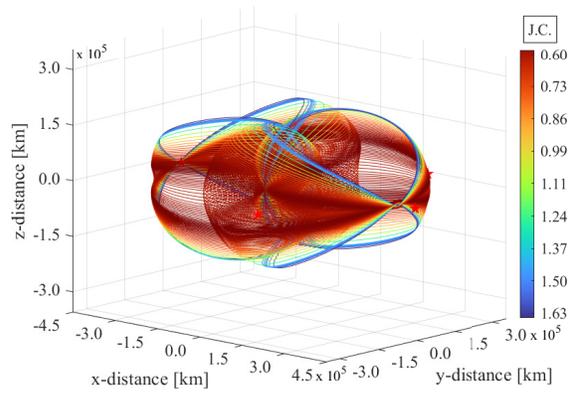

**(b) The retrograde-adjacent half: R2:1-S2**

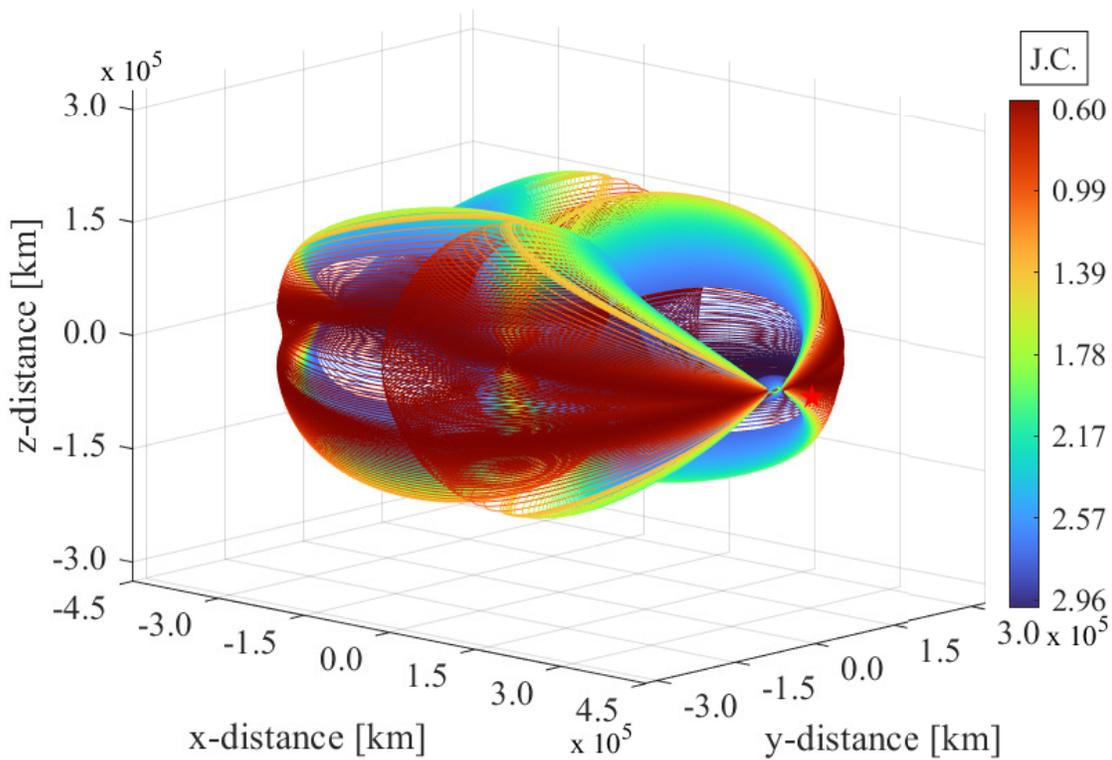

**Fig. 6   The 2:1 resonant spatial family (colloquially R2:1-S)**

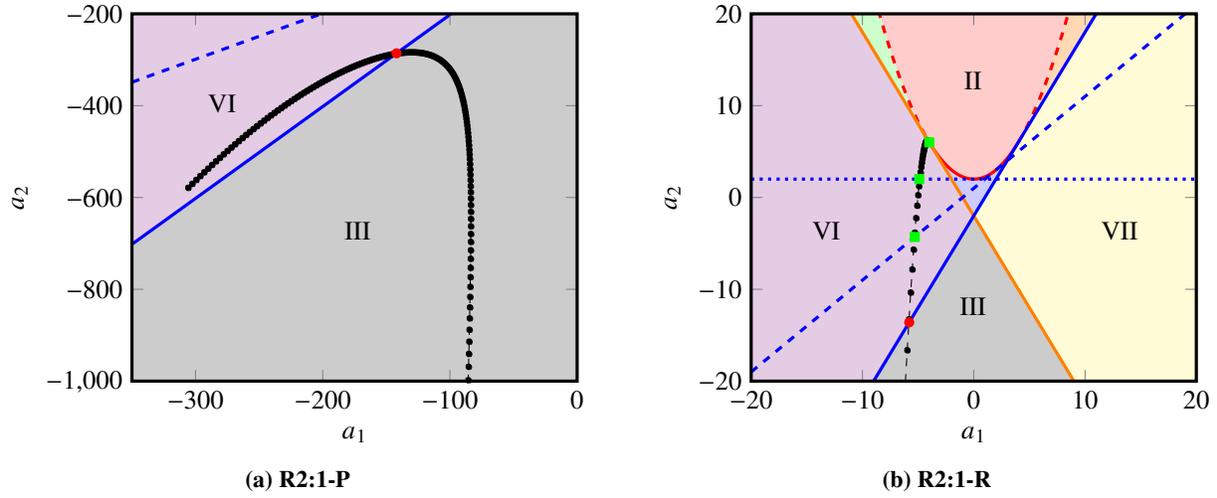

**(a) R2:1-P**

**(b) R2:1-R**

**Fig. 7   Broucke stability diagrams for planar R2:1-P and R2:1-R**

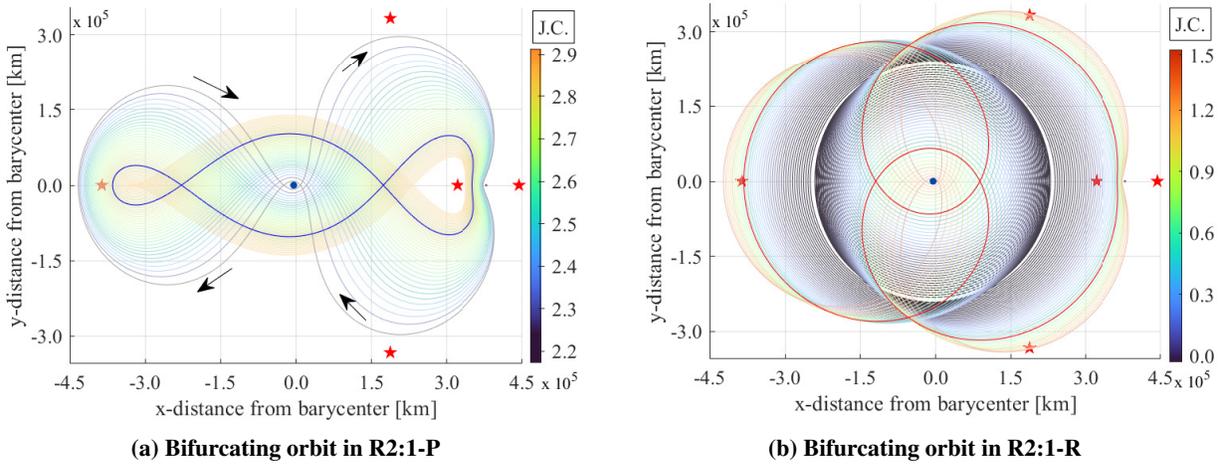

**(a) Bifurcating orbit in R2:1-P**

**(b) Bifurcating orbit in R2:1-R**

**Fig. 8   The period-doubling bifurcating orbits in R2:1-P and R2:1-R**

span a wide variety of high-energy *JC* values. Taking this fact into account, the data included in Figures 9 to 11 are all color-coded by their Jacobi constants. Doing so is intended to aid designers in orienting themselves within the family over the set of hodographs. Of note also is the range of *synodic periods* (or the rotating frame period of periodic states with respect to the position of the Moon) that members of R2:1-S posess. It is worth discussing the periods of each of these members in a nuanced way, as otherwise the analysis in this work may become unclear. To a first approximation, this paper began with the planar (and conic) 2:1 resonant orbit, a conic orbit that completed two revolutions for every one revolution of the Moon. After correction in the rotating frame, it appears that this trajectory becomes noticably different from a conic orbit (as seen in members of Figure 5). When correcting this conic trajectory in the CR3BP, we achieve a purely periodic trajectory in the rotating frame that is no longer conic (but still approximately so). When plotted in an inertial frame, the precise motion of this trajectory can be understood much more clearly. It retains its

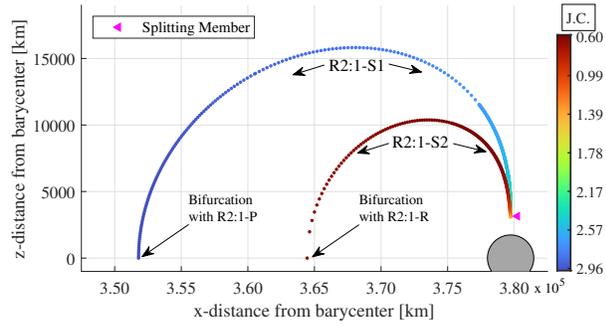

**(a) Position at Surface of Section $A$, x vs. z**

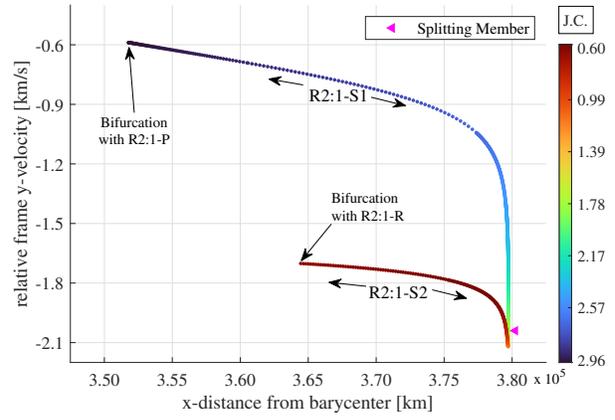

**(b) Relative velocity at surface of section $A$, x vs. $\dot{y}$**

**Fig. 9   Measuring crossings of R2:1-S across the surface of section**

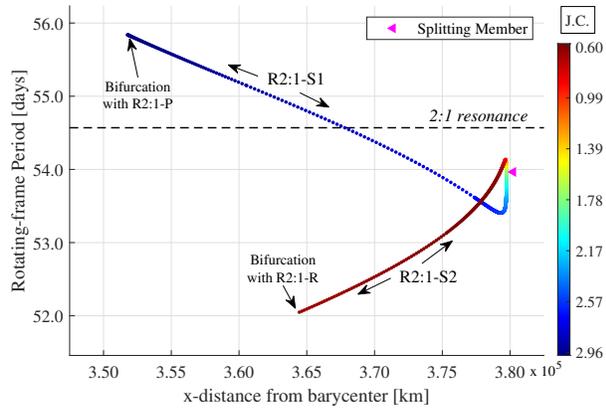

**(a) Period in the rotating frame, x vs. $\mathbb{P}$**

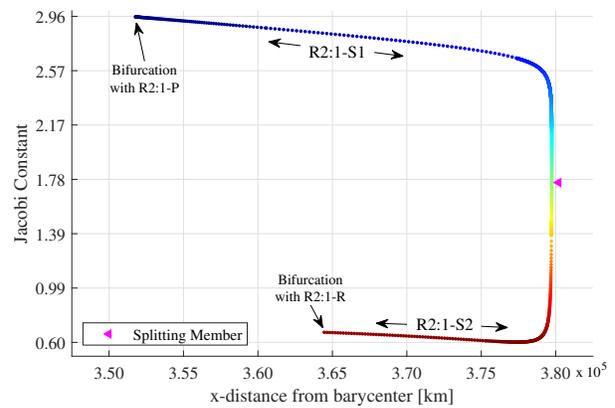

**(b) Jacobi constant for family members, x vs. $JC$**

**Fig. 10   Derived properties of R2:1-S, period and Jacobi Constant**

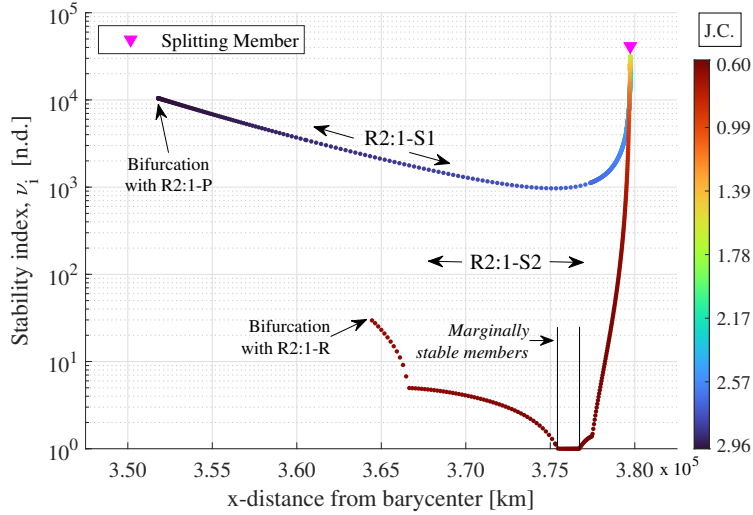

**Fig. 11    Stability index across the members of R2:1-S**

almost-conic behavior, completing approximately two almost-conic revolutions for every revolution of the Moon. It is worth noting that the synodic period of the members of Figure 5 generally is not exactly half of the sidereal period of the Moon's orbit. How to interpret this result? How can a rotating frame trajectory return to a precisely periodic configuration in less time than the Moon takes to orbit the Earth? Gaining insight into these questions can once again be done by examining the behavior in the inertial frame. These nearly-conic trajectories do indeed repeat in the rotating frame, but not so in the inertial frame. In the inertial frame, there is *apsidal precession* of the nearly-conic trajectories as the Moon's gravitational influence warps and shifts the trajectory. This apsidal precession rate is a direct function of the synodic period of the member. For nonsimple resonant motions, the *line of apsides* (in a conic sense) of the motion precesses about the $+\mathbf{u}_z$ axis on the xy plane in inertial space. For motions with a synodic period shorter than the lunar period, in the inertial frame, the line of apsides precesses at a negative rate about $+\mathbf{u}_z$. Looking at this precession from above the system, it appears that the line of apsides rotates in the xy plane in a clockwise manner. For motions with a synodic period longer than the lunar period, the opposite is true - in the inertial frame, the line of apsides precesses at a positive rate about $+\mathbf{u}_z$. Looking at this precession from above the system, it appears that the line of apsides rotates in the xy plane in a counterclockwise manner.

This behavior is retained in R2:1-S. *Apsidal precession* still occurs for members of R2:1-S and in the same way as it did for planar motions. The period-doubling bifurcations from which R2:1-S originate, in fact, double the synodic period required to achieve precise periodicity. These longer trajectories, in the inertial frame, still appear almost-conic, but now *inclined*. These inclined near-conics complete two revolutions before an encounter with the Moon, just as the planar ones do, but now the close approach with the Moon causes a change in inclination. This new trajectory is the mirror image of the original trajectory, inclined at the same angle magnitude with respect to the xy-plane, but inclined in the opposite direction. For example, when looking at a member of R2:1-S plotted inertially edge-on to the xy plane, the first two revolutions of the near-conic trajectory might appear to be inclined to the xy plane at 30 deg. After a close approach to the Moon, the near-conic trajectory would now appear to be inclined to the xy plane at −30 deg. Another pass by the Moon after another two subsequent revolutions restores the initial condition of the motion at an angle of 30 deg to the xy plane. These four revolutions about the Earth constitute one synodic period of a member of R2:1-S. Although the synodic period of a member of R2:1-S might be twice the period of the Moon around Earth, a satellite following this trajectory would still complete four revolutions of Earth and two lunar close approaches in the intervening time. It is for this reason that we still discuss these members as having a nearly 2:1 simple resonance with the Moon's orbit. Each satellite revolution around Earth still takes approximately half a month.

Of final note is Figure 11, which is a semi-log plot showing the stability index of each member of the family. Note the discontinuities and cusps in this graph[‡]. It is generally true that members of R2:1-S1 are less stable than members of R2:1-S2 in this sense. This more unstable behavior can be leveraged, specifically in trajectory design. Since R2:1-S1 members have a high stability index, departures and arrivals along these orbit's invariant manifolds are rapid. This behavior (described in Section V.A) is sought and used in Section VI. Not explored in this text but certainly of interest for future investigations is the fact that there are *linearly stable* members of R2:1-S2. These members occur relatively late in the family, but are spatial trajectories that exist in three dimensions with no stable or unstable subspace (in a linear sense). The efficacy of these orbits as *permanent parking orbits* existing between the Earth and Moon has not yet been explored, but is promising. This region of stability is indicated in Figure 11 under the label *"Marginally stable members"*.

# V. Design methodology for reaching cislunar destinations

## A. Relevant properties of the global manifolds of periodic orbits

In a nonlinear system of differential equations, stable and unstable global manifolds of a periodic orbit or fixed point $W^s$ and $W^u$ can be produced by propagating states selected from local eigenspaces $E^s$ and $E^u$ (respectively). These local (linear) eigenspaces are tangent to the global nonlinear manifolds at the periodic orbit or fixed point [20, Thm. 1.3.2]. In practice, selecting a state from the local eigenspace that approximates the nonlinear manifold at some fixed point or orbit is a challenge. We can, regardless, produce an estimate for a state in the manifold selected from the local eigenspace through a procedure detailed in Koon et al. [27, pg. 113] and reproduced here for stable and unstable manifolds. From a Floquet multiplier $\lambda \neq 0$, $\lambda \in \mathbb{R}$ and its associated eigenvector $\bar{D}(X_0) \in \mathbb{R}^6$ of the monodromy matrix $\mathcal{M}$ at the fixed point of an orbit $X_0$, it is possible to normalize the eigenvector by its first three components to create $D(X_0) = \frac{\bar{D}}{D_1^2 + D_2^2 + D_3^2}$. The estimated state in the local eigenspace is $X(X_0)$:

$$X(X_0) = X_0 \pm \varepsilon D(X_0) \tag{15}$$

---

[‡]There are a number of bifurcations that occur along the family which will remain unexplored in this text

The selection of $\varepsilon$ (colloquially called the 'step-off') can be done in a number of different ways [28] and is akin to more of an art than a science, but should be chosen to meet two competing conditions. First, the step-off should be small enough to approximate the local manifold well. Second, the stepoff should be large enough such that simulating trajectories from a stepoff are not subject to excessive numerical error as they diverge from the periodic orbit from which they are generated. For this paper, the value of $\varepsilon$ is chosen to give some physical significance to the step-off process. In particular, $\varepsilon$ is chosen to represent some dimensional distance in the problem (e.g. '25 km'). Multiplying this chosen $\varepsilon$ by $D$, this new state lies 25 km (in configuration space) from the fixed point of origin. Since $D \in \mathbb{R}^6$ (e.g. it has both a position and a velocity component), this 25 km step also adds some velocity deviation.

The global stable and unstable manifolds of a family of periodic orbits also display a 'nesting' behavior in that they cannot intersect in phase space. More precisely, two stable or two unstable manifolds of distinct fixed points or periodic orbits $X_1^*, X_2^*$ cannot intersect, nor can $W^s(X^*)$ (or $W^u(X^*)$) for some $X^*$ intersect itself [20, pg. 14]. Within the context of the CR3BP, this property of periodic orbit family members can be given a loose physical interpretation. Consider two adjacent members of the same family of periodic orbits. Each member has unstable and stable manifolds associated with it, and each manifold roughly shares that member's $JC$. Except in rare circumstances, adjacent members of the same family will have a different $JC$ value. Since they differ in energy, they will never intersect unless acted upon by some outside force that changes the $JC$ value (such as a thruster firing or orbital perturbations).

To visualize this discussion of nesting, we arbitrarily select two planar Lyapunov orbits about the Earth-Moon $L_1$ libration point and produce the globalized manifolds $W^u$ and $W^s$ associated with each orbit. In general, we represent stable ('attracting') manifolds using cooler colors, like light blue, and unstable ('repelling') manifolds using warmer colors, like light red. These two orbits (and their globalized manifolds) can be found in Figure 12 and Figure 13. Note that qualitatively, it seems that the manifolds of Figure 12 nest inside the manifolds of the larger Lyapunov visualized in Figure 13[§]. Qualitative features present in the smaller orbit's manifolds reappear in the larger orbit's manifolds but appear exaggerated (e.g. both family member's manifolds swing past the Earth in similar ways, but the larger orbit's manifolds pass closer to the Earth). Elsewhere, qualitative behavior changes in the manifolds of members along a family of orbits (i.e. manifolds of the larger orbit exiting the Earth-Moon system near the $L_2$ libration point, where the smaller orbit's manifolds do not). When considering a family of orbits for a design problem, this 'nesting' behavior allows a designer to, in a sense and within reason, 'tweak' where the manifolds of a design orbit reach in phase space (and when they arrive there) by choosing different family members. Since the 'nesting' behavior of manifolds between members of the same family of orbits is a general property of orbit families in dynamical systems, this principle can be leveraged despite the family of orbits considered. So long as the specific family member chosen is not mandated, by moving along a family of orbits, a designer has a degree of freedom not available when considering single orbits. In theory, this degree of freedom can be used to find manifolds that meet specific conditions.

---

[§]In fact, they do 'nest' when considering the orbits in phase space according to Guckenheimer and Holmes [20, pg. 14]

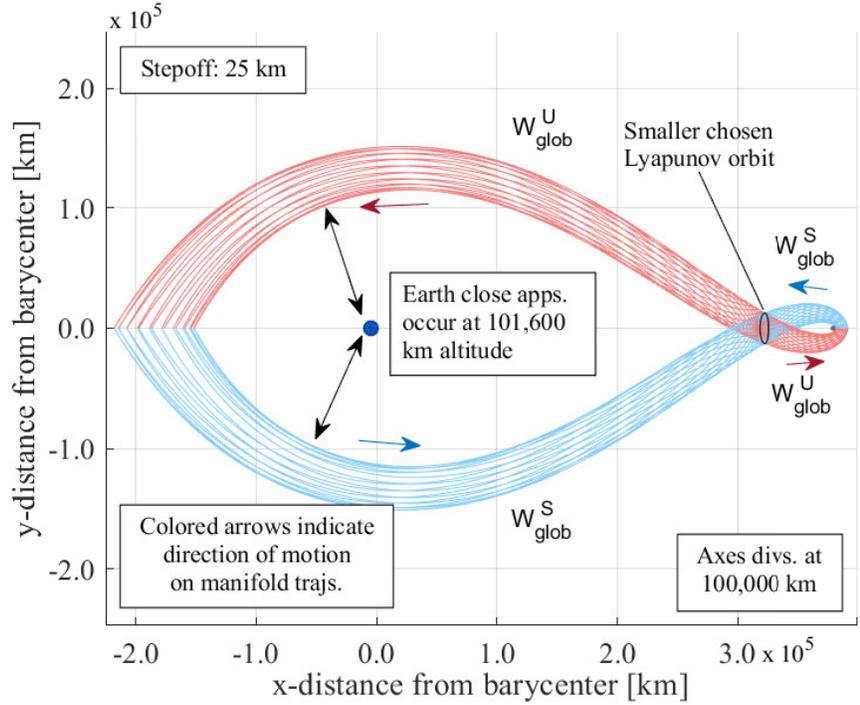

**Fig. 12   Global manifolds from the smaller member of the $L_1$ Lyapunov family**

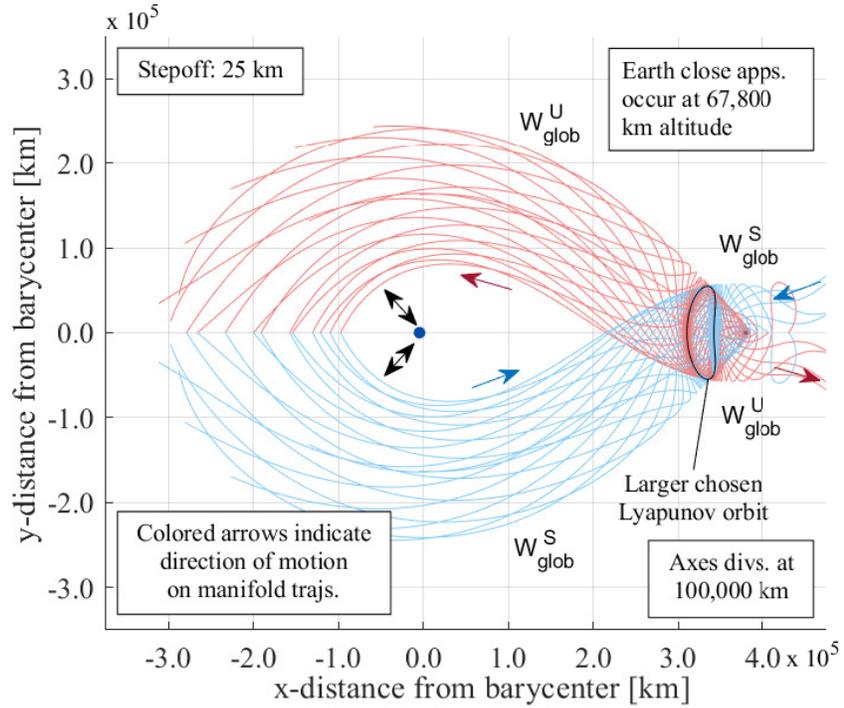

**Fig. 13   Global manifolds from the larger member of the $L_1$ Lyapunov family**

Furthermore, the full manifold of a periodic orbit does not need to be produced all at once. A manifold trajectory can be produced for *any* fixed point along an orbit and these fixed points can be sampled infinitely densely along an orbit (to within the limits of numerical precision). In fact, it is possible to produce a subset of a given manifold and analyze this section in isolation. Consider a periodic orbit with a period of $\mathbb{P}$ time units. If a trajectory designer selects some time along the orbit $\tau \in [0, \mathbb{P})$ (or some range of times along the orbit $[\tau_1, \tau_2]$), the manifold trajectories from that fixed point (or from within that range of fixed points) can be calculated and their manifolds can be constructed and analyzed. In tandem, this property and the nesting property described before can be used to iteratively target a desired end state (as long as that end state eventually lies within some manifold of some orbit in the family). This technique is used to target a position intersection with the halo orbits in Section VI.

As can be seen in Figure 14, the globalized stable and unstable manifolds of a single periodic orbit also have some symmetries. Note that this discussion is only relevant for CR3BP periodic orbit families that are symmetric about the xz plane (such as R2:1-P, R2:1-R, R2:1-S the Lyapunovs, the halos, and others). A Lyapunov orbit is again selected for the sake of demonstration, and its globalized manifolds are plotted until their circumlunar crossing with the xz-plane. We return again to the Mirror Theorem [21] (restated here as applied to the CR3BP) to describe the mechanism by which these symmetries arise for orbits symmetrical about the xz-plane. According to the Mirror Theorem, these orbits and the Earth/Moon must each have positions and velocities perpendicular to eachother at the same instant. In this configuration, some symmetries are guaranteed. In the CR3BP, this statement is equivalent to asserting that the trajectory of $m_3$ cross the xz plane in a perpendicular fashion. Having two such perpendicular crossings ensures periodicity in the CR3BP. Others are guaranteed by symmetries within the differential equations. For this class of orbits, there are four symmetries in the manifolds and, generally, in the trajectories of the CR3BP. The symmetries are presented with the property of $W^u(X^*)$ on the left and the symmetric property of $W^s(X^*)$ on the right:

1) $\tau \leftrightarrow -\tau$

2) $y \leftrightarrow -y$

3) $\dot{x} \leftrightarrow -\dot{x}$

4) $\dot{z} \leftrightarrow -\dot{z}$

All other elements of the state are otherwise identical between the two manifolds. These symmetries are visualized specifically in Figure 14 but can be seen in any diagram that visualizes the stable and unstable manifolds of a Mirror Theorem periodic orbit. Additionally, in diagrams with only one manifold of a Mirror Theorem periodic orbit plotted, the symmetries should be inferred by default. Note that in the CR3BP, all the above statements are *exactly* true, but in other models with perturbative terms or in non-autonomous models (e.g. ephemeris), these relationships will only *approximately* hold and only for moderately short time horizons. The relationships will still hold qualitatively in many of these more complicated cases, can be used to gain insight into the more complicated flow, and can be used as a starting point for trajectory design.

Additionally, these symmetries in time and space allow a designer to design for one scenario and somewhat generalize the design results to other scenarios. For example, a designer could reason about relationships between stable and unstable manifolds when looking for homoclinic connections from and to a Mirror Theorem orbit. Consider the case where the global unstable manifold of this periodic orbit is computed, and the designer seeks to find a homoclinic connection back to this original orbit. This homoclinic connection occurs at locations where the stable and unstable manifolds of one periodic orbit intersect simultaneously in position and velocity space. Because of the symmetries discussed above, the designer need not produce the stable manifold as well to search for these connections. The designer can simply look for locations where the unstable manifold crosses the xz-plane in a perpendicular manner. In this perpendicular crossing scenario, the $x-$ and $z-$ velocities on the unstable manifold are zero at the crossing. Since this occurs at a crossing of the xz plane, $y = 0$. Under this condition, the flow from the unstable manifold of the orbit will naturally flow directly into the stable manifold of the same orbit, achieving the desired homoclinic connection.

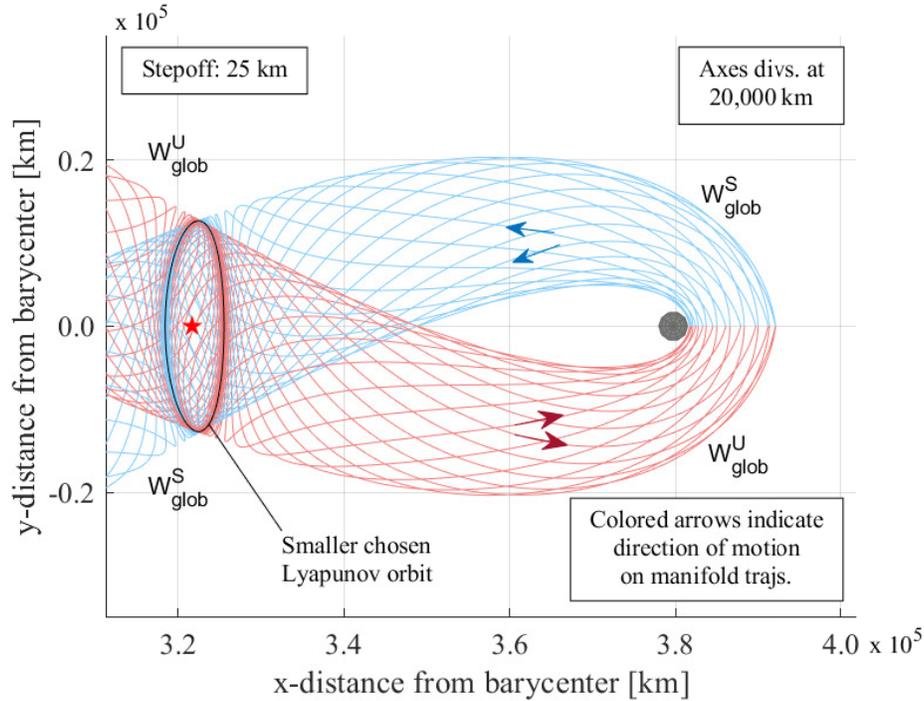

**Fig. 14   Stable and Unstable Manifold Time and Space Symmetries**

This idea of searching for perpendicular crossings of manifolds is also relevant in cases where the crossing is not perfectly perpendicular. For example, by searching for near-perpendicular crossings, a designer can identify opportunities with low fuel cost for recovery into the original periodic orbit. Consider again the case where a designer intends to design a trajectory from the unstable manifold of an orbit onto the stable manifold of the same orbit. If a near-perpendicular crossing is identified, the designer can place a maneuver prior to that crossing to modify the trajectory. The goal of this maneuver should be to modify the trajectory so that the arrival at the crossing occurs in a truly perpendicular manner. If done far enough ahead of time, a designer can leverage the Mirror Theorem [21] to reinsert into the original periodic orbit inexpensively. Denote the time between the designed maneuver and the crossing of the transfer trajectory with the xz plane as $T$. Due to the Mirror Theorem, performing such a maneuver is guaranteed to cause the transfer trajectory to intersect with the stable manifold in configuration space $2T$ time units after the maneuver is performed. With a second maneuver of identical size performed at $2T$, a designer can reinsert onto the stable manifold of the original periodic orbit. This is exactly the process used in the next section to find cheap reinsertion maneuvers in Figures 21 and 22.

## B. Use of momentum exchange tethers and relevant physics

Momentum exchange tethers (METs) have been discussed and their dynamics extensively studied in the literature [29]. It is not the goal of this article to expand on this research, but merely to describe how METs are an interesting use case for members of R2:1-S. To this end, METs are long cables connected to heavy counterweights capable of delivering $\Delta v$'s to payloads through catch, velocity vector redirection and release of the payload [30]. Due to the conservation of angular momentum in steady-state rotation, this structure spins about an angular momentum vector fixed in inertial space. Catching and throwing payloads is a largely conservative action, allowing angular momentum in the spinning tether to be traded for propulsive effort with or against the direction of spin. Pairing such tethers allows for a trade of angular momentum between them, potentially affecting large $\Delta v$s that retain the total system angular momentum and cost little propellant. Moving in a simple circular motion and roughly modeled as a rigid body, the tip of the tether travels at an instantaneous and constant velocity $v_{\text{tip}}$. It has been shown that METs can deliver any $\Delta v$ within a solid disk of $\Delta v$ with radius $2v_{\text{tip}}$, that is, $\|\Delta v\| \leq 2v_{\text{tip}}$ [31]. It can deliver this $\Delta v$ in cases where the starting and ending velocities relative to the tether's translational motion have the same magnitude. Where they have different magnitudes, the tether can shorten or lengthen, leveraging the conservation of angular momentum to speed up or slow down the payload during the redirection phase. In inertial space, this disk of possible (or 'producible') $\Delta v$ vectors lies in the plane normal to the angular momentum vector of the MET. In the rotating frame, the disk of $\Delta v$ is more difficult to describe. The disk precesses with the frame, much like a spun coin settling on a table. This precession introduces variability in $\Delta v$ that is producible, but this variability may be undesirable based on the design case. Specifically, the variability introduces a nonautonomous component into the design problem, leading to complications when designing in an autonomous *native design model*. To negate this precession and simplify the design process, it is helpful to align the MET's angular momentum vector with the $\pm z$-direction. Doing so unfortunately precludes the MET's ability to deliver out-of-plane $\Delta v$ (that is, $\Delta v$ with a $z$-component). However, this restriction is not an issue - a trajectory designer simply needs to find locations where METs can deliver $\Delta v$ in the plane, as is done in Section VI.

Furthermore, with a naive transfer design, one tether may throw a payload to another tether linking those two tethers by the payload's trajectory arc in space. Qualitatively, this transfer causes the payload to leave one tether and arrive at another at a specified time. What a naive approach does not account for is what happens if the destination tether fails to catch the payload or, for some other reason, aborts. Using a periodic trajectory such as a member of R2:1-S as the basis for a transfer design allows for a baked-in bias towards safety. When considering the natural flow in a transfer design, designers can be confident that operational factors will not affect mission safety. For example, METs may miss a catch or need to abort a catch. The payload spiraling into interplanetary space after a missed catch is suboptimal. It is desirable for that payload to be guaranteed not to collide with the Moon and return to the vicinity of Earth (or back into some periodic trajectory) for recovery. In our case, it is precisely a member of R2:1-S that can help achieve this goal.

## VI. Case study: traversal to a member of the $L_1$ halo family

In this paper, R2:1-S has been described many times as capable of 'reaching' a versatile set of locations that span cislunar space. The mechanism by which the orbits 'reach' destinations is via the use of their stable and unstable manifolds. Using the concepts described in Section V.A, a trajectory designer can scan across members of a periodic orbit family and simultaneously through various values $\tau$, the position along a given member in time. The trajectory designer can then find manifold trajectories or segments that satisfy the desired end states, provided that the desired end states $\mathbf{X} \in W^u$ or $\mathbf{X} \in W^s$. In this case study, we consider a design problem in which a designer seeks to link the regime below Geostationary Orbit (GEO) with a multibody orbit near the Moon using METs. Taking on the mantle of the designer, we place two METs: one in a subgeostationary circular orbit and the other in a multibody orbit near the Moon. This is done with the hypothetical goal of designing an infrastructure capable of sending supplies and receiving trash from the Lunar Gateway station from Earth orbit for little to no round trip $\Delta v$ cost. Arbitrarily at the cislunar end, we choose to target a member of the Northern $L_1$ halo family. This halo was chosen to have a similar $JC$ to the $L_2$ 9:2 lunar sidereal-resonant halo under the assumption that cheap transfers (or free heteroclinic connections) between each are possible [32]. As this is not the objective of this article, these heteroclinic connections or their feasibility have not been investigated. It should be noted that the sidereal resonant and the synodic resonant (when measured with respect to the phases of the Moon) $L_2$ halo have different but not entirely dissimilar $JC$ values. Therefore, we specifically seek to place a MET in a sub-GEO orbit, a MET in the Northern $L_1$ halo of our choice, and seek to use members of R2:1-S to accomplish the transfer between these two. In an ideal case, the transfer would not cost fuel beyond what is required for stationkeeping in the selected member of R2:1-S and all impulsive maneuvers are performed by METs (subject to the constraints described in Section V.B). A layout of the design problem is presented in Figure 15.

With this mission concept in mind, the ideas presented throughout the paper are used here to achieve two goals in this case study. First, find a member of R2:1-S whose stable manifold 1) passes near Earth (ideally near GEO, 2) has a small time of flight between this close approach and stepoff, and 3) has a near-zero $z$ direction velocity at close approach. Second, to find a member of R2:1-S with unstable manifolds that 1) intersect our target $L_1$ halo, 2) has a small time of flight between stepoff and this intersection, and 3) matches the halo's $z$-direction velocity at the intersection. Fortunately, there is a single member of R2:1-S1 that meets all the prescribed conditions, as visualized in Figure 16. This member was chosen from R2:1-S1 due to the stability properties (or more precisely, the *instability*) of

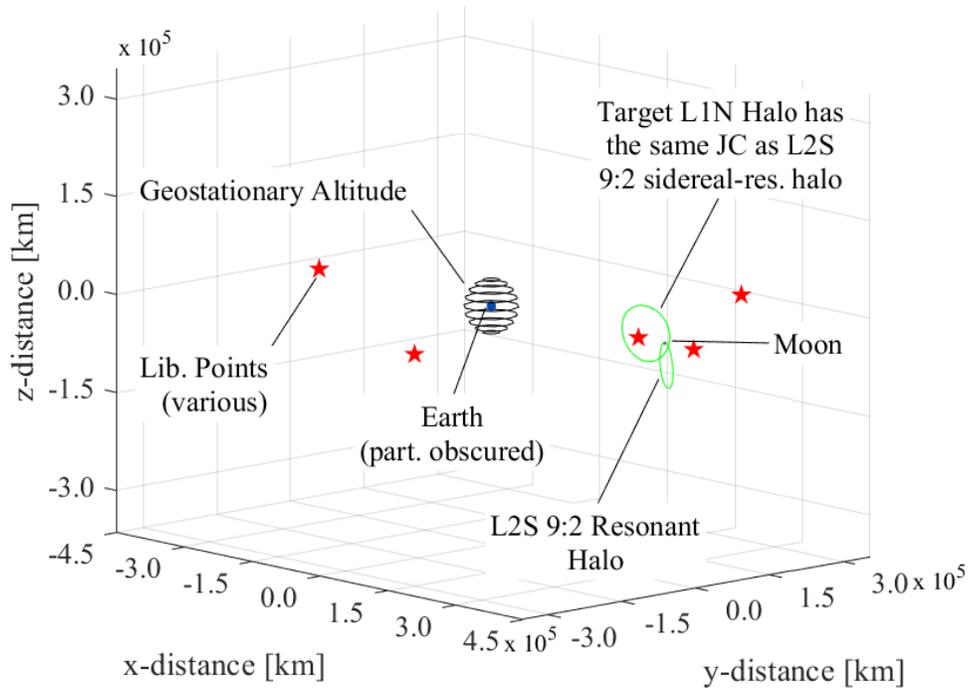

**Fig. 15    Design Schematic for $L_1$ halo-bound payload case study**

the half-family. Two views of the member are presented in Figure 16, first a projective view onto the plane and second a

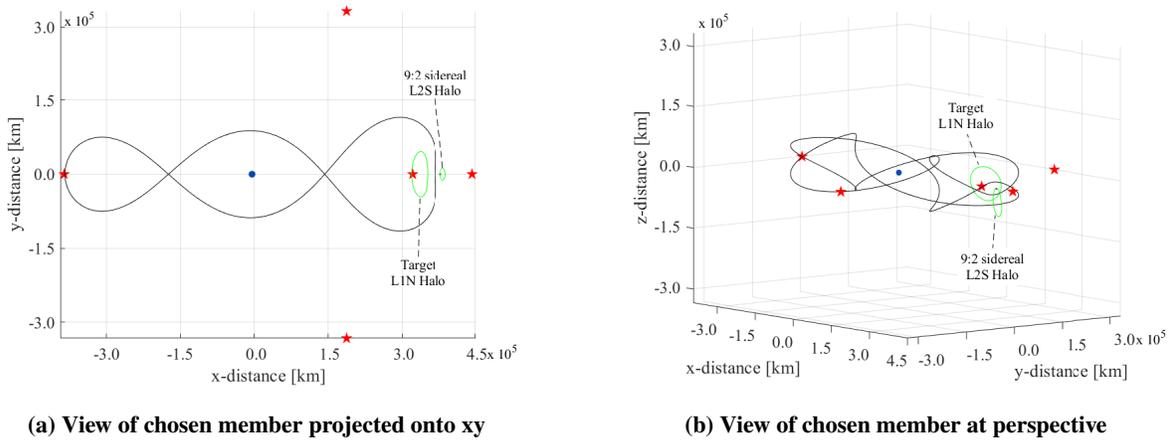

**(a) View of chosen member projected onto xy**

**(b) View of chosen member at perspective**

**Fig. 16    One member of R2:1-S1 (plotted in black) satisfies all constraints**

view of the three-dimensional structure of the orbit. For the remainder of this text, spatial diagrams like Figure 16(b) will be presented as projected onto the xy plane, but three-dimensional behavior is still present. Where appropriate, the z-component of points in space or of velocities will be reported in writing. The member has the orbital properties detailed in Table 2:

**Table 2    Initial conditions, properties of selected member of R2:1-S1**

| $x_0$ [km] | $z_0$ [km] | $\dot{y}_0$ [km/s] | $JC$ [n.d.] | $\mathbb{P}_{syn}$ [days] | Stab. Index $\nu_i$ [n.d.] |
|---|---|---|---|---|---|
| 368966 | 15789 | -0.82796 | 2.78959 | 54.47 | 1418.34 |

## A. Reaching below Geostationary Orbit using Stable Manifolds

We start by addressing the first goal. When examining the stable manifolds of the chosen member, the subset of stable manifold trajectories that passes below GEO is identified. Furthermore, a specific trajectory below this altitude is sought that has a $\mathbf{u}_z$-velocity of nearly 0 m/s at close approach for the reasons outlined in Section V.B. One such stable manifold close approach trajectory is identified and displayed in Figure 17. The identified perigee occurs at an Earth altitude of 27109 km (2819.5 km above the xy plane) at a speed of 4.861 km/s ($-0.0094$ km/s in the $\mathbf{u}_z$ direction). The flight time along the manifold trajectory between the close approach and the eventual step-off condition is approximately one synodic period of the chosen member, or exactly 58.2 days. Injection into the chosen member of R2:1-S1 can be done more quickly than 58.2 days with an additional maneuver or set of maneuvers at an additional fuel cost.

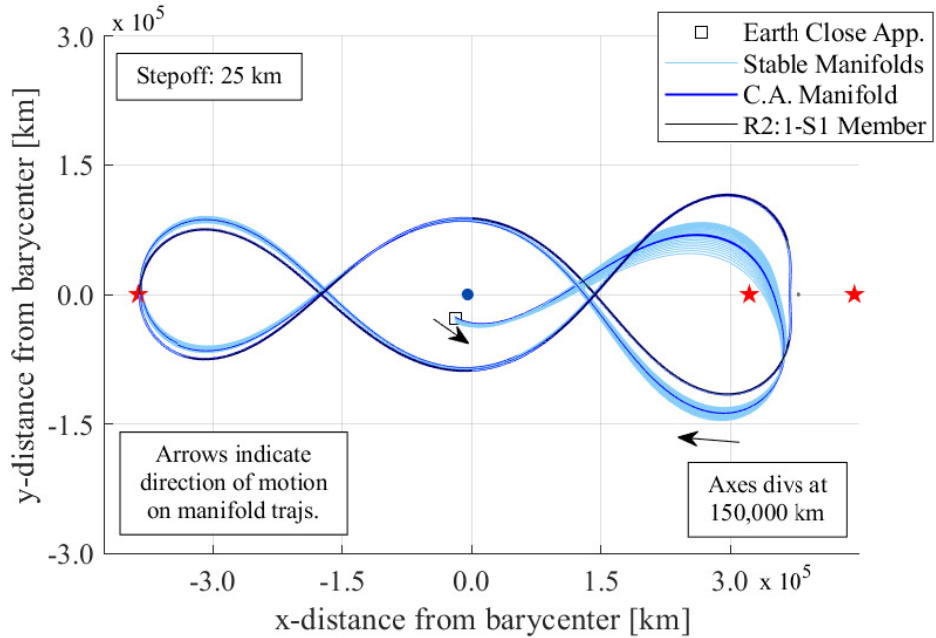

**Fig. 17    Stable manifolds of selected R2:1-S1 member reach below GEO**

Of interest is the circular orbit around Earth (at 27109 km altitude) tangent to this close approach. In particular, a MET placed in the circular tangential orbit at this altitude would have a speed of 3.45 km/s in it's orbit. This speed (and others to follow) is calculated using the Keplerian two-body conic model. The speeds of the identified close approach and the MET are colinear and differ in magnitude by 1.411 km/s. Note that to reach this tangential circular orbit from Earth, we will need a Geostationary Transfer Orbit (GTO)-like Hohmann transfer from LEO. Presuming that a typical GTO-like transfer has an apogee at the circular altitude and a periapsis at 200 km altitude, the speed at apogee on this transfer orbit is 1.977 km/s. The speed difference between the transfer orbit and the MET is 1.473 km/s. Placing a MET in this circular orbit effectively splits the two speeds (close approach speed and the speed at the apogee of the transfer orbit) and, according to the physics described in Section V.B, is an ideal placement to allow the MET to produce the required total $\Delta v$. As such, the total $\Delta v$ required is within the limits of previously designed METs [30] and is found by adding the two speed differences. This $\Delta v$ requirement is 1.411 + 1.473 = 2.884 km/s, which implies a tip speed of at least half this number. The tip speed of the tether is required on average to be greater than 1.442 km/s, which again lies within the limits of the previously designed METs.

**B. Reaching the Chosen $L_1$ Halo Orbit using Unstable Manifolds**

For the same member, we can also address the second goal shown in Figure 18. Using a section of the unstable manifold of our chosen member, we are able to find a range of locations where the unstable manifold intersects with the chosen halo in position space. Along this set of intersections, we searched for intersections whose velocities in the $\mathbf{u}_z$ direction matched. After a cursory search through the manifold, one manifold trajectory was found to intersect the halo at a roughly similar velocity in the $\mathbf{u}_z$ direction. This is the opportunity to catch visualized in Figure 18. At the catch opportunity, the unstable manifold and the halo intersect in position space. This catch opportunity occurs after stepoff at a time of flight of approximately one synodic period of the chosen member, or exactly 61.12 days after the step-off condition and at $-25793$ km out of the plane. This out-of-plane amplitude is approximately half of the maximal out-of-plane amplitude of the chosen member of R2:1-S1 (roughly $50,000$ km for this chosen member). Similarly to what was stated in the previous subsection, the departure from our R2:1-S1 member can be done more quickly with an additional maneuver or a set of maneuvers at an additional fuel cost.

When comparing the manifold velocity with the halo velocity, the velocities differ by a magnitude of 0.3935 km/s (of which 0.0062 km/s is out of the plane). Performing a $\Delta v$ of this magnitude here would result in injection into the halo. This is by no means the manifold with a planar maneuver that minimizes this in-plane $\Delta v$ amplitude, but for our design case study it is optimal enough. After catch, the MET would need to throw the payload somewhere to send it to its final lunar destination. Preliminary results indicate that a throw into another member of the $L_1$ halos is possible via that other member's stable manifolds. This throw, at a magnitude of 0.472 km/s, is done so that the payload lands on the stable manifold of a separate member of the Southern variant of the $L_1$ halo family. After leaving the MET, the manifolds of this new (and unstable) Southern halo can be used to find connections with other multibody orbits or cislunar destinations. As before, a planar $\Delta v$ is amenable to using METs to accomplish the $\Delta v$ and is considerably smaller than the $\Delta v$ values asked of previous MET designs. For this cislunar halo, a tip speed of 434 m/s on average is appropriate. In fact, parking a MET in this $L_1$ halo would allow the use of this MET to transfer payloads to the orbit of the LOP-G space station via heteroclinic transfer or low-cost transfer [33]. The design of a MET with such a comparably undemanding tip speed seems convincingly compatible with modern day materials and launch capabilities, but until further validation this claim remains speculative.

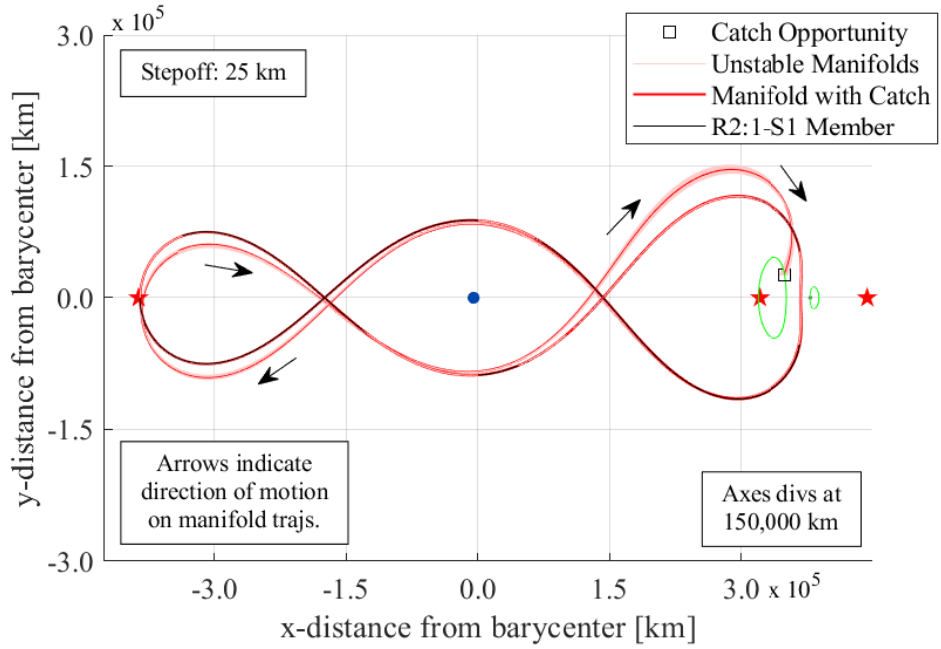

**Fig. 18  Unstable manifolds of selected R2:1-S1 member provide catch opportunity at the chosen $L_1$ halo**

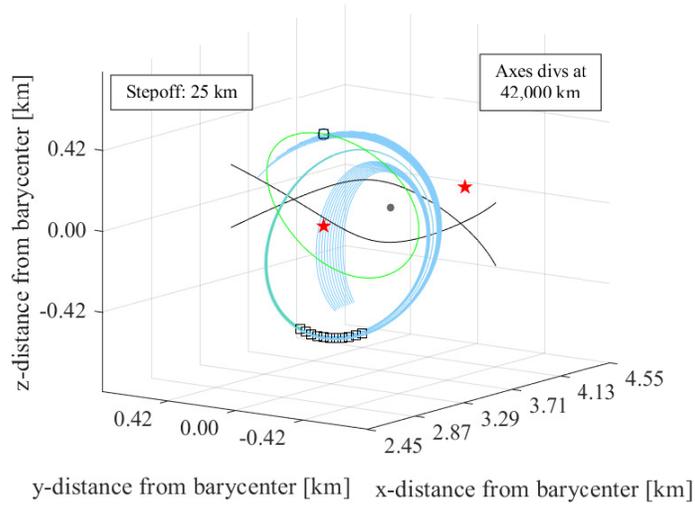

**Fig. 19  The cislunar tether can throw directly to other multibody orbits**

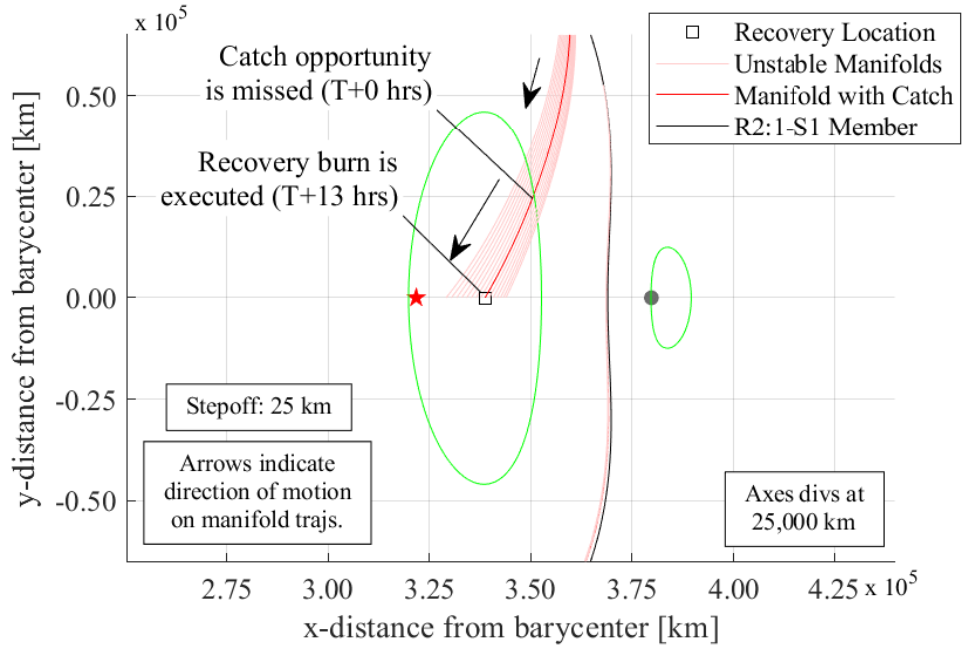

**Fig. 20  Top-down view of an aborted catch procedure showing trajectory continuing to the xz-plane**

## C. Responding to a Missed or Aborted Catch

Consider a hypothetical situation where the cislunar MET misses its catch. We assume that planning an adequate recovery action may take time, and implementing any recovery actions occurs T + 13 hours after missed catch. This attempted catch, miss, and recovery period is visualized in Figure 20. In the vein of recovering from a missed catch, it is possible to find *nearly* homoclinic connections back to our R2:1-S design member. These nearly homoclinic connections contain multiple low-altitude Earth passes each. Each recovery is implemented with two recovery impulses, each performed at the prescribed recovery location and separated in time. For each recovery discussed, an impulsive maneuver of size $||\Delta v||$ is first performed at the recovery location at $T + 13$ hours. The dynamics after the impulse transports the beleaguered spacecraft back to the recovery location after a period of time, denoted $T_r$. Along the way, close approaches to Earth deliver opportunities for a return to the Earth's surface at typical cislunar return speeds within a few days. If Earth abort options are not taken, an additional and final impulsive maneuver (also of size $||\Delta v||$) inserts the spacecraft into the stable manifold of the original periodic orbit. For the sake of this simplified strategy, stable manifold insertion ends the crisis and returns the spacecraft to a safe configuration to resume normal operations. We demonstrate two such recoveries: in the first case (Figure 21) we find a total recovery time of $T_r = 26.3$ days at a cost of $2 \cdot ||\Delta v|| = 141$ m/s. In the second case (Figure 22), we find a slower recovery with a recovery time of $T_r = 109.1$ days but at a much lower cost of $2 \cdot ||\Delta v|| = 43$ m/s. These recoveries are in no way $\Delta v$ optimal, but still show that members of R2:1-S are amenable to safe mission design under reasonable operational assumptions.

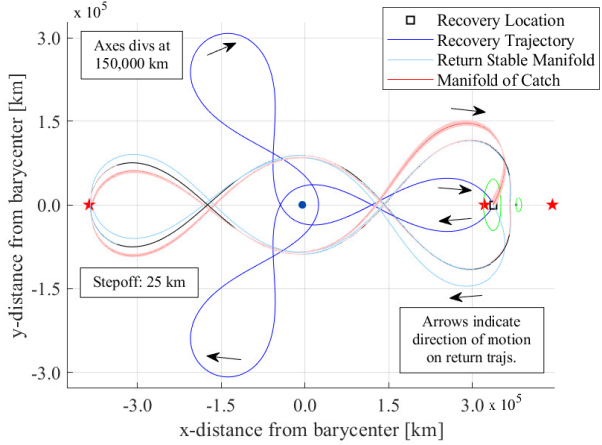

**(a) Total recovery trajectory**

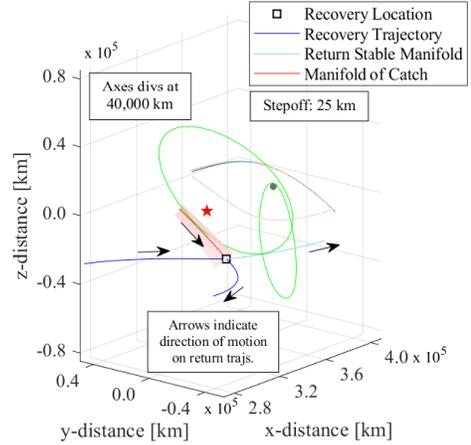

**(b) Recovery location, zoomed**

**Fig. 21 Cheap reconnections into the chosen R2:1-S1 member are feasible: Option 1**

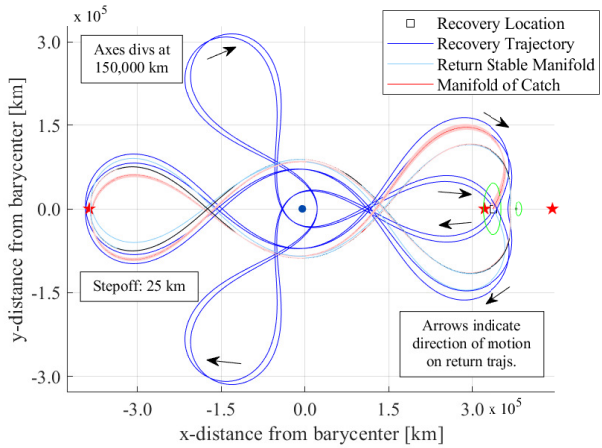

**(a) Total recovery trajectory**

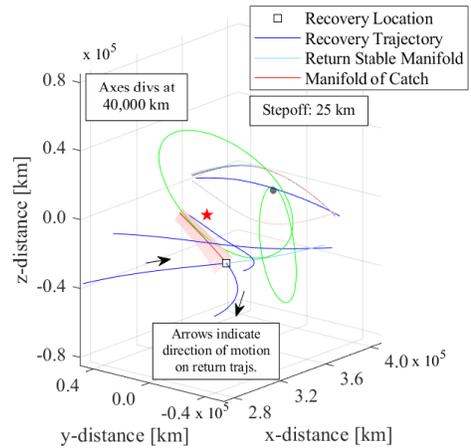

**(b) Recovery location, zoomed**

**Fig. 22 Cheap reconnections into the chosen R2:1-S1 member are feasible: Option 2**

It should be noted that all the designs in Section VI are only point solutions to the design problem. While interesting in its own right, each solution provided here should foremost be seen as proof that these solution classes exist more generally. No designs shown here came from an overly constrained process with precisely only one solution. Furthermore, hypothetical steps to optimize these designs seem straightforward. With further research, it may be possible to find other point solutions or perhaps families of point solutions for MET pair systems in the Earth-Moon system. Building on previous works such as Byrnes et al. [34], it may even be possible to find multibody solutions between Earth and other planets in the solar system.

# VII. Conclusion

This work focuses on the generation and study of resonant orbit families in the Circular Restricted Three-Body Problem (CR3BP). Specifically, this paper identifies a gap in the literature among the currently present families of three-dimensional resonant orbits and seeks to fill that gap. To this end, this paper introduces the 2:1 resonant spatial family (abbreviated R2:1-S) in the spatial CR3BP as a generalization of the previous planar 2:1 resonant prograde family and the 2:1 resonant retrograde family (abbreviated R2:1-P and R2:1-R, respectively) previously discussed in the literature. R2:1-S is shown to emerge from each family through a period-doubling bifurcation and is shown to continuously link the two families in phase space. Using a number of numerical tools, R2:1-S is analyzed to produce family hodographs and to understand the stability properties of family members. Of particular interest, the family can be split into two at the most unstable member (in a Lyapunov sense), and each half is shown to share qualitative properties. In the second half of the family, referred to as R2:1-S2, it is shown that there are marginally stable members.

Furthermore, this work shows that this resonant orbit family is capable of meeting a set of conditions set out at the beginning of this paper. A family of orbits that spans three-dimensional space, passes near Earth and the Moon, and repeats this dynamic at a frequent rate was sought. R2:1-S meets these requirements. Using these properties, a hypothetical case study is performed to demonstrate the efficacy of these orbits as a mission design tool. The problem of designing a reusable cislunar infrastructure system using momentum exchange tethers (MET) is presented as an interesting application to showcase complex mission design. In this study, globalized manifolds of family members are explored and shown to reach various near-Earth and cislunar locales. Furthermore, they are proven to have relationships that are conducive to the use of METs with other multibody orbits. As an interesting result of the constrained design, each MET is separately implementable and each is shown to lie well within the design limits derived in other works. The cislunar tether, at first glance, is even shown to be feasible assuming no technological developments.

Regardless of content, future work may develop the R2:1-S family to a degree that it is more easily usable in future missions. If this is the case and developments continue, periodic return trajectories that exist in three-dimensions may be the key to transforming cislunar space and the Lunar environment from being a place to simply visit into a place to stay.